\documentclass[12pt]{article}

\usepackage{jheppub}

\usepackage{amsmath,amsfonts,amsthm,bm,bbold}
\usepackage[T1]{fontenc}

\usepackage{booktabs}
\usepackage{placeins}
\usepackage{mymacros}
\usepackage{makecell}

\usepackage{pgfplots}
\pgfplotsset{compat=1.18}

\newcommand\tra{^{\mathpalette\raiseT\intercal}}

\newcommand\raiseT[2]{%
\setbox0\hbox{$#1{#2}$}\raise\dp0\box0}

\makeatletter
\@addtoreset{equation}{section}
\makeatother

\renewcommand{\vec}{\mathbf}

\renewcommand{\t}{\theta}
\newcommand{\m}{\mu}

\newcommand{\eps}{\epsilon}

\begin{document}

\title{How natural are Froggatt--Nielsen models?}
\author[a]{Gero von Gersdorff}
\author[a]{and Lucas  Modesto}
\affiliation[a]{Pontifícia Universidade Católica do Rio de Janeiro\\
Rua Marquês de São Vicente 225, Rio de Janeiro, Brazil}
\emailAdd{gersdorff@puc-rio.br}
\emailAdd{lucasmodesto@aluno.puc-rio.br}

\abstract{
The Froggatt--Nielsen (FN) mechanism aims at a natural explanation of the surprising hierarchical structure of the masses and mixings of  Standard Model quarks and leptons.
We test this claimed naturalness against various known and some original measures of fine-tuning, including various local measures as well as Bayes factors. The FN models serve as a fruitful laboratory for testing and comparing these measures.
We find that the standard Barbieri-Giudice metric is overestimating naturalness in many cases, as it does not sufficiently account for the failure of a model to be realized at order-one parameter values, even though the latter are
favoured by an appropriate prior distribution.
As a side product of our study, we present simple geometric expressions for the posterior distributions of the parameters on the manifold of best-fit values.
}

\maketitle

\section{Introduction}

The very hierarchical structure of fermion masses and mixings is one of the unresolved mysteries of the Standard Model (SM).  Disregarding the neutrino sector, the properties of the fermions are encoded in their Yukawa couplings, complex three by three matrices $Y_u$, $Y_d$ and $Y_e$.
In particular, the singular values of these matrices are proportional to the masses, and the unitary Cabibbo-Kobayashi-Maskawa (CKM) matrix appearing in the charged current interactions of the quarks is a product of the basis rotations of the left handed quark fields that diagonalize the Yukawa couplings $Y_u$ and $Y_d$.
Both the quark masses and the CKM matrix elements display a strong hierarchy as summarized in table \ref{tab:data}.

\begin{table}[h]
\centering
\small
\begin{tabular*}{\textwidth}{@{\extracolsep{\fill}}c c c c@{}}
\toprule
Up-type Yukawas & Down-type Yukawas & CKM parameters & Charged-lepton Yukawas \\
\midrule
\begin{tabular}[t]{@{}r@{\hspace{0.45em}}l@{}}
$y_u$ & $6.3\times10^{-6}$ \\
$y_c$ & $3.1\times10^{-3}$ \\
$y_t$ & $0.87$
\end{tabular}
&
\begin{tabular}[t]{@{}r@{\hspace{0.45em}}l@{}}
$y_d$ & $1.4\times10^{-5}$ \\
$y_s$ & $2.7\times10^{-4}$ \\
$y_b$ & $1.4\times10^{-2}$
\end{tabular}
&
\begin{tabular}[t]{@{}r@{\hspace{0.45em}}l@{}}
$\theta_{12}$ & $0.23$ \\
$\theta_{23}$ & $4.2\times10^{-2}$ \\
$\theta_{13}$ & $3.7\times10^{-3}$ \\
$\delta_{13}$ & $1.2$
\end{tabular}
&
\begin{tabular}[t]{@{}r@{\hspace{0.45em}}l@{}}
$y_e$ & $2.8\times10^{-6}$ \\
$y_\mu$ & $6.0\times10^{-4}$ \\
$y_\tau$ & $1.0\times10^{-2}$
\end{tabular} \\
\bottomrule
\end{tabular*}
\caption{Yukawa couplings and CKM parameters at $1\,\mathrm{TeV}$
in the SM $\overline{\mathrm{MS}}$ scheme, from ref.~\cite{Antusch:2013jca}.
We use the standard CKM
parametrization~\cite{ParticleDataGroup:2024cfk}, the phase $\delta_{13}$ is
given in radians.}
\label{tab:data}
\end{table}

Over the years, many approaches have been explored to account for this unusual pattern, including Abelian
 \cite{Froggatt:1978nt}  and non-Abelian \cite{Pakvasa:1977in,Wilczek:1977uh,Barbieri:1995uv,Barbieri:1997tu} flavor symmetries,
 Yukawa textures \cite{Weinberg:1977hb,Fritzsch:1977za},
extra dimensions \cite{Grossman:1999ra,Gherghetta:2000qt,Huber:2000ie,Cabrer:2011qb,vonGersdorff:2013rwa},
and clockwork models \cite{Burdman:2012sb,vonGersdorff:2017iym,Patel:2017pct,Alonso:2018bcg,Smolkovic:2019jow,AbreudeSouza:2019ixc,vonGersdorff:2020ods}.

Most approaches have in common that a natural explanation is sought by aiming for a structure
\be
(Y_u)_{ij}=\eps_{q^i}\,(\hat Y_{u})_{ij}\,\eps_{u^j}\,,
\qquad
(Y_d)_{ij}=\eps_{q^i}\,(\hat Y_{d})_{ij}\,\eps_{d^j}\,,
\qquad
(Y_e)_{ij}=\eps_{\ell^i}\,(\hat Y_{e})_{ij}\,\eps_{e^j}\,,
\label{eq:flavorgen}
\ee
where the $\eps_{f^i}$ are real numbers that are naturally hierarchical  while the proto-Yukawa matrices $\hat Y_f$ are order one couplings. Notice that for the quark sector, the same  matrix
$\diag(\eps_{q_1},\eps_{q_2},\eps_{q_3})$ multiplies both $Y_u$ and $Y_d$ from  the left.
Within an SU$(5)$ grand unification (GUT) framework, the $\eps_f$ matrices typically respect SU$(5)$,
\be
\eps_q=\eps_u=\eps_e\,,\qquad \eps_d=\eps_\ell\,,
\ee
whereas some SU$(5)$ breaking effect is introduced via $\hat Y_d\neq \hat Y_e\tra$.

Depending on the model, there may be further constraints on the $\epsilon$ parameters. For instance, in FN models,  they are of the form
\be
(Y_u)_{ij}=\eps^{Q_{q_i}-Q_{u_j}}\,(\hat Y_{u})_{ij}\,,
\qquad
(Y_d)_{ij}=\eps^{Q_{q_i}-Q_{d_j}}\,(\hat Y_{d})_{ij}\,,
\qquad
(Y_e)_{ij}=\eps^{Q_{\ell_i}-Q_{e_j}}\,(\hat Y_{e})_{ij}\,,
\label{eq:FNcharges}
\ee
where the $Q_{f^i}$ are integer charges, and $\eps$ is a parameter that is of the order of $0.1$.
Here we take the convention that the charges of the left handed fields are positive and those of the right handed fields are negative.
Models may also constrain the order-one matrices $\hat Y_{u,d,e}$, for instance, by setting to zero some of their components or imposing relations between them, typically dictated by some
symmetry relation.

For the purpose of this work, we will commit to the FN parametrization eq.~(\ref{eq:FNcharges}), even though the analyses will be very similar for e.g. extra-dimensional or partial-compositeness models.
The reason the FN parametrization works is that in the singular value decomposition (SVD)
\be
\begin{aligned}
Y_u={}&U_{u_L}\diag(y_u,y_c,y_t)\, U_{u_R}^\dagger\,,\\
Y_d={}&U_{d_L}\diag(y_d,y_s,y_b)\, U_{d_R}^\dagger\,,\\
Y_e={}&U_{e_L}\diag(y_e,y_\mu,y_\tau)\, U_{e_R}^\dagger\,,
\end{aligned}
\ee
the change-of-basis matrices are already close to diagonal, in the sense that (up to order-one numbers)
\be
(U_{u_L})_{ij}\sim \eps^{|Q_{q_i}-Q_{q_j}|}\,,\qquad (U_{u_R})_{ij}\sim \eps^{|Q_{u_i}-Q_{u_j}|}\,,
\ee
etc, and hence the singular values scale as
\be
y_{u_i}\sim \eps^{Q_{q_i}-Q_{u_i}}\,,
\qquad
y_{d_i}\sim \eps^{Q_{q_i}-Q_{d_i}}\,,
\qquad
y_{e_i}\sim \eps^{Q_{\ell_i}-Q_{e_i}}\,,
\label{eq:svsFN}
\ee
whereas the CKM matrix satisfies
\be
(V_{\rm CKM})_{ij}\sim \eps^{|Q_{q_i}-Q_{q_j}|}\,.
\label{eq:ckmFN}
\ee
These results are very well known of course. For generalizations to arbitrary flavor models as well as a systematic correction to the leading terms, as well as exact bounds on their errors see refs.~\cite{vonGersdorff:2019gle,vonGersdorff:2025zox}.

Statistical approaches to Froggatt--Nielsen models have studied how readily random order-one coefficients reproduce the observed flavor hierarchies, both for lepton flavor \cite{Altarelli:2002sg,Altarelli:2012ia,Aloni:2021wzk,Bergstrom:2014owa} and in systematic searches over quark charge assignments \cite{Cornella:2023zme}. More recently, ref.~\cite{Ibe:2024cvi} performed a comprehensive Bayesian comparison of FN textures using Bayesian evidences obtained by marginalizing over their continuous parameters. Building  on this probabilistic viewpoint, in this paper we compare Bayesian evidences against traditional local
fine-tuning metrics averaged over the posterior.
The results allow us to compare different FN textures to one another and identify particularly natural ones. We also compute the naturalness metrics for the SM proper and quantify the improvement
achieved by the FN models. The flavor models thus serve as a environment to test and compare the various metrics of naturalness, identifying strengths and weaknesses of each of them.

The remainder of this paper is organized as follows: sec.~\ref{sec:posterior} develops the posterior on the best-fit manifold, sec.~\ref{sec:naturalness-metrics} introduces our naturalness measures, sec.~\ref{sec:results} presents results for the SM and FN textures, and sec.~\ref{sec:discussion} discusses their implications.  The appendices collect details on treatment of complex parameters and the numerical implementation.

\section{The  posterior distribution}
\label{sec:posterior}

\subsection{General models}

Often the number of parameters $d_P$ exceeds the number of observables $d_O$. There are therefore many choices that reproduce the
observed data within their errors, we will refer to these choices as {\em realizations} of the model.
The best fit values  define a hypersurface in parameter space that we will dub the best-fit manifold.

Since a model with more parameters than observables has infinitely many realizations, we  typically want to study ensembles of them.
Often in the flavor literature, these ensembles aim to reproduce physical observables only up to $\mathcal O(1)$ numbers, hoping that the derived conclusions
(such as predictions for flavor changing neutral currents) do not
depend too much on this rough approximation.
Instead, in this work, we will construct ensembles of true realizations.
The natural context for this is the Bayesian posterior for the parameters of a given model,
\be
p(\t|\mu_{\rm exp})\sim L(\t)p(\t)\,,
\label{eq:post}
\ee
where $p(\t)$ is some prior density that models our expectation for the parameters before taking data into account. The other quantity on the right hand side,
\be
L(\t)= \exp(-\chi^2(\t)/2)\,,\qquad \chi^2(\theta)=\sum_n \left(\frac{\m^n(\t)-\mu^n_{\text{exp}}}{\sigma^{n}_{\text{exp}}}\right)^2\,,
\ee
is the likelihood function. The posterior describes the relative plausibility of viable parameter points after conditioning the prior parameter distribution on the observed data.
The best fit manifold is the solution of the equations  $\m^n(\t)-\m^n_{\rm exp}=0$.

The posterior probability has a very neat form in the "thin-transversal-limit", in which we formally take the experimental uncertainties to zero.
 In this case, the support of the
posterior collapses to the best fit manifold:
\be
 e^{-\chi^2(\theta)/2}p(\t)d{\bm \t}
	\quad \to\quad \delta(\m(\theta)-\m_{\rm exp})	p(\t)d{\bm \t}\,,
\ee
up to an irrelevant normalization factor.\footnote{Unlike the likelihood,
the delta function is in fact a normalized distribution for the data for given parameters, $p(\mu_{\rm exp}|\theta)=\delta(\m(\theta)-\mu_{\rm exp})$ in the thin limit.}
Whenever we are interested in averages of quantities that do not vary rapidly in the transversal directions we may neglect the transversal thickness of the posterior and work in the thin limit.
We thus  change variables $\t\to(v,u)$, where $v$ are normal directions and $u$ are tangential ones.
Then, by the coarea formula, the (unnormalized) posterior can be written as
\be
	 \delta(\m(\theta)-\m_{\rm exp})	p(\t)d{\bm \t}
	=\sqrt{\frac{\det g(u)}{\det h(u)}}p(\t(u))\delta(v)d\vec v d\vec u\,.
\ee
Here
\be
g_{ij}(u)\equiv \frac{\partial\theta^a}{\partial {u^i}}\frac{\partial\theta^a}{\partial u^j}
\ee
is the induced metric on the best fit manifold $M$ (with respect to the Euclidean one in the $\t$ variables), and
\be
h^{kl}(u)\equiv \frac{\partial\m^k}{\partial {\t^a}}\frac{\partial\m^l}{\partial \t^a}\bigg|_{\theta(u)}
\ee
 is the inverse of the normal metric on the manifold.

Integrating over normal directions  gives the posterior distribution on the manifold $M$, (up to an $u$-independent normalization constant)
\be
p(u|\mu_{\rm exp})d\vec u\sim \sqrt{\frac{\det g(u)}{\det h(u)}}p(\t(u))d\vec u\,.
\ee
We recognize the induced volume form $\sqrt{\det g}\,d\vec u$ in the numerator.
Notice that under a reparametrization of the manifold $M$, the measure $p(u|\mu_{\rm exp})d\vec u$ stays invariant ($p$ and $\sqrt{\det h}$ are scalar functions on $M$).

\subsection{Flavor models}

So far these have been general considerations. We now specialize to the case of the SM flavor sector and the FN models, which will lead to further simplifications.

We define a flavor model by the Yukawa parametrizations eq.~(\ref{eq:flavorgen}) with a fixed set of $\eps_{f^i}$ parameters.
Furthermore, we focus on the FN parametrization eq.~(\ref{eq:FNcharges})
 and restrict our attention to the quark sector only.
The lepton sector is completely decoupled from the quark sector (unless we consider GUT scenarios).
Every FN model is thus uniquely fixed by its {\em texture} $\{\eps,Q_{q_i},Q_{u_i},Q_{d_i}\}$.
The free parameters within each model are then the proto-Yukawa couplings $\hat Y_u$ and $\hat Y_d$.
For quark flavor models with $d_P=36$ (real) parameters and $d_O=10$ observables this defines a 26 dimensional best fit manifold.

 In the SM, the  $\theta^i$ are  the  Yukawa couplings $Y_u$ and $Y_d$, these are 36 real parameters.
The 26 dimensional manifold $M$ is precisely the 26 dimensional flavor group
\be
M=\frac{ U(3)_q\times U(3)_u\times U(3)_d}{U(1)_B}\,,
\label{eq:MSM}
\ee
where the three $U(3)$ factors are the unitary rotations acting on the $q_L$, $u_R$ and $d_R$ fields respectively, and $U(1)_B$ is the diagonal $U(1)$ subgroup (i.e., Baryon number).
Indeed, the embedding map into the full parameter space $M\hookrightarrow P$ can be given explicitly as
\footnote{Any representative of the quotient eq.~(\ref{eq:MSM}) yields the same Yukawa couplings, so that the map is well defined on $M$.}
\be
\begin{aligned}
Y_u(u)={}&U_q(u)\diag(y_u,y_c,y_t)U_u^\dagger(u)\,, \\
Y_d(u)={}&U_q(u)V_{\rm CKM}\diag(y_d,y_s,y_b)U_d^\dagger(u)\,.
\end{aligned}
\label{eq:SMyukawaorbit}
\ee
The flat metric $\sum_a (d\theta^a)^2=\tr dY_u^{}dY_u^\dagger+\tr dY_d^{}dY_d^\dagger$ is invariant under the flavor group.
Since the embedding of $M$ into parameter space is equivariant, its pullback $g$ is an invariant metric on the homogeneous space $M=U(3)^3/U(1)_B$.
Its volume form is therefore, up to an overall normalization, the invariant measure induced by the Haar measure on $U(3)^3$.
We denote this measure by $dH(u)$.
Equivalently, it can be generated by drawing $U_q$, $U_u$ and $U_d$ independently from the Haar measure on $U(3)$, where their common $U(1)_B$ transformation acts trivially.
The flavor group also acts transitively on $M$, while leaving the observables  invariant.
As a consequence, the normal metric $h^{kl}$ is  constant on $M$.
Finally, a Gaussian prior
\be
p\sim e^{-\frac{1}{2\sigma^2}(\tr Y_u^{}Y_u^\dagger+\tr Y_d^{}Y_d^\dagger)}
\ee
is clearly invariant under the flavor group and hence a constant on $M$.
Therefore, the Bayesian posterior for the SM and a Gaussian prior  is simply
\be
p(u|\mu_{\rm exp})d\vec u\sim \sqrt{\det g(u)}d\vec u \equiv dH(u)\,,
\ee
i.e., the invariant measure induced on the quotient.

Next, we move on to FN models. For now we treat only the proto-Yukawa couplings stochastically, and keep the texture, i.e., the charge assignment and the value of $\eps$ as given.  The proto-Yukawas are  linear rescalings of the SM Yukawa couplings, see eq.~(\ref{eq:flavorgen}), inducing a simple embedding
\be
\hat Y^{}_x(u)=E_q^{-1}Y_x^{}(u)E_x^{-1}\,,
\label{eq:embed}
\ee
where
\be
E_q\equiv\diag(\eps^{Q_{q_1}},\eps^{Q_{q_2}},\eps^{Q_{q_3}})\,,\qquad
E_x\equiv\diag(\eps^{|Q_{x_1}|},\eps^{|Q_{x_2}|},\eps^{|Q_{x_3}|})\,.
\ee
Such a linear relation  factorizes out of the ratio of the Jacobians in the coarea formula
\footnote{The easiest way to see this is by making a change of variables from $\hat Y$ to $Y$ before applying the coarea formula. Without using the coarea formula, the identity  eq.~(\ref{eq:constantrescaling}) remains quite nontrivial. It relies on the orthogonality of the  rows of $\partial \mu/\partial \theta$ and the columns of $\partial\theta/\partial u$.}
\be
\frac{\sqrt{\det g_{\rm FN}(u)}}{\sqrt{\det h_{\rm FN}(u)}}=\eps^{-Q_{\rm tot}} \frac{\sqrt{\det g_{\rm SM}(u)}}{\sqrt{\det h_{\rm SM}}}\,,
\label{eq:constantrescaling}
\ee
where $g_{\rm FN}$ and $h_{\rm FN}$ are based on $\hat Y(u)$, and $g_{\rm SM}$ and $h_{\rm SM}$ on $Y(u)$, and
\be
Q_{\rm tot}=2\sum_{i}( 6 Q_{q_i}-3Q_{u_i}-3Q_{d_i})\,.
\ee
Therefore, in the FN case with a  Gaussian prior, the posterior on the best-fit manifold becomes
\be
p_{Q,\eps}(u|\mu_{\rm exp})d\vec u\sim e^{-V_{Q,\eps}(u)} dH(u)\,,
\label{eq:FNposterior}
\ee
where
\be
V_{Q,\eps}(u)=\frac{\tr \hat Y_u^{}(u)\hat Y_u^\dagger(u)+\tr \hat Y_d^{}(u)\hat Y_d^\dagger(u) }{2\sigma^2}\,,
\label{eq:FNpotential}
\ee
where $Q$ is the charge assignment.

\section{Naturalness metrics}

\label{eq:measures}
\label{sec:naturalness-metrics}

Intuitively we consider a model fully natural if two conditions are met.
Firstly, the physical values of dimensionless Lagrangian parameters should not be anomalously small or large.
Secondly,  the sensitivity of observables to small variations of
these parameters should be at most of order one.

The first condition suggests
drawing the parameters from suitable priors that strongly suppress very small or very large parameters.
The actual physical values of the parameters are then obtained by turning this prior into a posterior, as described in section \ref{sec:posterior}. In general it is however not guaranteed that the
posterior is centered around order one parameters.

The second condition requires some kind of local sensitivity measure.
The most famous of such metrics is the
Barbieri--Giudice matrix \cite{Barbieri:1987fn}
\be
(S_{\rm log})^i_{\ j}=\frac{\partial\log \mu^i}{\partial \log\t^j}\,.
\label{eq:BG}
\ee
It is a local measure for the sensitivity of the observable with respect to the parameters. For instance,
$|(S_{\rm log})^i_{\ j}|=10$ means that a, say, 10\% change in $\theta_j$ induces a 100\% change in $\mu_i$.
If the parameters are dimensionless (as is the case in the present work), one might also consider a linear instead of a log derivative
\be
(S_{\rm lin})^i_{\ j}=\frac{\partial\log \m^i}{\partial  \t^j}\,.
\label{eq:BG2}
\ee
The different choices are related to how we are measuring distances in parameter space.
The linear prescription implies that we are using the metric  $ds_{\rm lin}^2=\sum_i (d\theta^i)^2$,
while the logarithmic one corresponds to $ds^2_{\rm log}=\sum_i (d\log \t^i)^2$.
For $O(1)$ parameters the logarithmic and linear metrics give similar results.

Whatever the choice, it is very suggestive to use the spectrum of  singular values of the matrix $S$. Since $d_P>d_O$,  there are $\operatorname{rank} S=d_O$ singular values.
We define the following metrics:
\begin{itemize}
\item
Maximum Singular Value (MSV). This corresponds to the most sensitive direction of the matrix $S$.
\item
Product of Singular Values (PSV). This is equivalent to \be J\equiv\sqrt{\det S^{}S\tra}\ee Comparing to MSV, this quantity identifies fine-tuning that is not concentrated in one particular direction.
This measure of tuning was proposed previously in ref.~\cite{Fichet:2012sn}.
\end{itemize}
Moreover, the singular vectors contain further useful information. The dominant left singular vector of $S$ points in the direction of observable space which has the highest sensitivity to varying the input parameters, and the
dominant right singular vector points in the corresponding direction of parameter space.
If we are not interested in the latter, we can consider the eigenvalues and eigenvectors of the matrices $S^{}S\tra$, which contain the scalar products (with respect to the parameter space metric $ds^2$) of the parameter space gradients $\bm \nabla_\t \log\mu^n$.

The local sensitivity metrics generally detect pathological (unnatural) points, but  may give both false positives and false negatives.\footnote{We adopt here the usual medical terminology where negative is good and positive is bad.} The former arise for instance when  a model contains an inherently sensitive function whose purpose is to turn an approximately order one parameter into a very large or small observable.
Famous examples include the proton mass as a function of the QCD coupling, $m_p=\Lambda_{\rm UV} e^{-c/g^2}$. In the context of the FN model, one such sensitivity is given by the $\epsilon$ parameter, even though
the sensitivities to it are power law and not exponential as in the QCD example.
Anderson and Castaño \cite{Anderson:1994dz} suggested comparing the physical parameter region to the generic parameter region in order to assess whether such a sensitivity is "worse than average".
Here we will simply ignore the sensitivity of observables to variations of $\epsilon$, and only include the derivatives with respect to the proto-Yukawa couplings.

A false negative on the other hand can occur if sensitivity metrics are of order-one but occur at very small or large parameters. As an example in our context, consider a SM at the point where $Y_u$ is diagonal. Then the sensitivity of
$\partial\log(y_u)/\partial\log(Y_u)_{11}=1$ but $(Y_u)_{11}\sim 10^{-6}$.
It is worth emphasizing that this kind of false negative may persist after averaging despite the use of priors that emphasize order-one parameters. This happens if the posterior is very different from the prior and selects very unlikely a priori parameter regions.
One way to detect such false negatives is to realize that the log and linear metrics are complementary in that regard, since $(S_{\rm log})_{ij}=(S_{\rm lin})_{ij}\theta^j$. Therefore, a discrepancy between these metrics indicates that low values in one of them is due to anomalously large or small parameters.
Two more robust diagnostics will be developed below.
Finally, the PSV may give a false negative for a different reason. SVs of, say $s\sim 0.01$ should be regarded as no more natural than $s\sim 1$, but $s\sim 1$ are more natural than $s\sim 100$.  The PSV metric then has the obvious problem that possible SVs $\ll 1$ will distort the overall metric. Therefore, an $O(1)$ PSV should be cross checked against other measures, or one should simply consider the full spectrum of SVs.

In the rest of this section we describe two alternatives to address the false negatives of the sensitivity measures that better capture the criterion of order-one couplings. The first possibility is to use a prior-aware metric on parameter space. To this end, assume that the parameters are independently
distributed and define new coordinates via $p^i(\t^i)d\t^i=dq^i(\t^i)$, with  $0\leq q^i\leq 1$. Measuring distances in parameter space with the Euclidean metric of the $q^i$ coordinates gives
\be
\frac{\partial \log \mu^i}{\partial q^j}
=(S_{\rm lin})^i_{\ j}\frac{1}{p^j(\t^j)}\,.
\ee
Unlikely parameter points therefore receive a penalty inversely proportional to their prior probability density. We further calibrate these coordinates by insisting that for $|\t^i|\sim 1$,
the metric agrees with the Euclidean metric in the $\t$ coordinates. This is equivalent to the constant rescaling $\tilde q^i(\t^i)=q^i(\t^i)/p^i(1)$ and using the flat metric in the $\tilde q^i$ coordinates. The calibrated sensitivity matrix used below is therefore
\be
(S_{\rm puc})^i_{\ j}\equiv\frac{\partial\log\mu^i}{\partial\tilde q^j}
=(S_{\rm lin})^i_{\ j}\frac{p^j(1)}{p^j(\t^j)}\,.
\ee
These coordinates have  the property that any prior becomes uniform, we thus refer to them as prior-uniform coordinates (PUCs). It
furthermore has the property that the volume form derived from the metric $ds^2_{\rm puc}$ is, up to a normalization constant, the prior measure $p(\t)d{\bm \t}$ itself.

The second possibility is to use Bayesian evidence \cite{Cabrera_2009,Fichet:2012sn}.
Bayes' theorem relates posterior and prior as
\be
p(\t|\mu_{\rm exp})=\frac{p(\mu_{\rm exp}|\t)p(\t)}{p(\mu_{\rm exp})}\,.
\ee
Here $p(\mu_{\rm exp})$ is called the model's evidence, i.e., the probability density of the data given the model. Integrating Bayes' theorem over the parameter space, we can write
\be
p(\mu_{\rm exp})=\int_P p(\mu_{\rm exp}|\t)p(\t)d\bm\t\,.
\label{eq:evidence}
\ee
The evidence measures how well a model as a whole explains the data. In particular, if the likelihood forces the valid parameter points to lie in a region of low prior probability,
the evidence is reduced. Furthermore,  the evidence  can detect cancellations of parameters. This is because cancellations naturally occupy only a small, non-generic volume of parameter space and hence tend to be suppressed under reasonable priors.
Therefore, the evidence serves as a useful metric for naturalness  \cite{Fichet:2012sn}.
Since $p(\mu_{\rm exp})$ is a density, it is more meaningful when taking ratios of evidences between models.
These ratios are called
 Bayes factors (BFs) between two models for the same data $\mu_{\rm exp}$.
\be
B_{1,2}\equiv \frac{p_1(\mu_{\rm exp})}{p_2(\mu_{\rm exp})}\,.
\ee
Explicitly, in terms of the likelihoods
\be
B_{1,2}
=
\frac{\int_{P_1} L_1(\t)p_1(\t)d\bm\t}{\int_{P_2} L_2(\t)p_2(\t)d\bm\t}\,,
\ee
where $P_i$ are the parameter spaces of each model, $L_i(\t)=\exp(-\chi_i^2(\t)/2)$ are the likelihoods, and $p_i$ are the normalized priors.
The BF  fully takes into account the
$\mathcal O(1)$ condition via the prior,  but is only a relative measure between two models.
Another caveat concerns the right choice of prior. The latter must accurately reflect our  order-one condition. For instance, a unit Gaussian has $P(|\t|>5)\sim 10^{-6}$, driving the evidence to zero
at such parameter values. If one considers $|\t|\sim 5$ still natural, the uniform Gaussian prior produces false positives for the evidence.
Needless to say, the prior choice is not an objective matter, and the desired prior-sensitivity inevitably comes with a strong prior-dependence.

There is also an interesting connection between evidence and the prior-uniform PSV metric.
As already pointed out in ref.~\cite{Fichet:2012sn},  the PSV metric in any coordinates appears naturally in the denominator of the integrand of eq.~(\ref{eq:evidence}). In  PUCs, the prior itself is absorbed:
from eq.~(\ref{eq:evidence}), using $p(\mu_{\rm exp}|\t)=\delta(\mu(\t)-\mu_{\rm exp})$, and applying the coarea formula in the  prior-uniform coordinates, gives for the evidence
\be
p(\mu_{\rm exp})= \int_M \frac{\sqrt{\det g_{\rm puc}(u)}}{\sqrt{\det h_{\rm puc}(u)}}du\,.
\label{eq:observation}
\ee
The denominator is, up to a model and  $u$ independent constant, nothing  but the prior uniform PSV metric $J_{\rm puc}$
\be
\sqrt{\det h_{\rm puc}(u)}=(\prod_i \mu_{\rm exp}^i) J_{\rm puc}(u)\,.
\ee
 Therefore, models with  large $J_{\rm puc}$ tend to suppress the evidence.
 Of course, there is still the numerator that does depend on $u$ and hence can impact the BFs.
However, we will see that the BF metric and the $J_{\rm puc}$ metric will produce naturalness rankings of FN models that are highly correlated.

In the case of the FN models, using the thin limit of the likelihood and the manipulations in section \ref{sec:posterior} (in particular, eq.~(\ref{eq:constantrescaling})), the Bayes factor between
two FN textures can be written as
\be
B_{(Q,\eps),(Q',\eps')}
=\frac{ \eps^{-Q_{\rm tot}} \int_M dH(u)e^{-V_{Q,\eps}(u)}}{ \eps'^{-Q'_{\rm tot}} \int_M dH(u)e^{-V_{Q',\eps'}(u)}}\,.
\ee
Furthermore, by introducing a prior $p(\eps)$ for the $\eps$ parameter, we can also compare evidences of charge assignments
\be
B_{Q,Q'}
=\frac{\int d\eps\, p(\eps) \eps^{-Q_{\rm tot}} \int_M dH(u)e^{-V_{Q,\eps}(u)}}{\int d\eps\, p(\eps) \eps^{-Q'_{\rm tot}} \int_M dH(u)e^{-V_{Q',\eps}(u)}}\,.
\label{eq:BQQ}
\ee
Eq.~(\ref{eq:BQQ}) will be the quantity reported in section \ref{sec:FN}.
Finally, the BF between a given FN  model with charge assignment $Q$   and the SM can be given explicitly as
\be
B_{Q, \text{SM}}
=\frac{\int d\eps\, p(\eps) \eps^{-Q_{\rm tot}} \int_M dH(u)e^{-V_{Q,\eps}(u)+V_{\rm SM}}}{\int_M dH(u)}\,,
\label{eq:FN-SM-BF}
\ee
where
\be
\begin{aligned}
V_{\rm SM}={}&\frac{1}{2\sigma^2}\sum_q y_q^2\approx \frac{y_t^2}{2\sigma^2}\,.
\end{aligned}
\ee
Notice that even though $e^{-V_{Q,\eps}+V_{\rm SM}}$ is tiny for order-one proto Yukawa couplings, the Jacobian factor $\eps^{-Q_{\rm tot}}$ by far outweighs this suppression.

\section{Results}
\label{sec:results}

The fact that $\mathcal O(1)$ dimensionless couplings are considered natural should be implemented by the priors. We will employ a Gaussian prior with mean zero and variance one for both the real and imaginary parts of the Yukawa matrices, i.e.,
\be
p(\hat Y)\sim
e^{-\frac{1}{2}\tr\!\left(\hat Y_u^{}\hat Y_u^\dagger
+\hat Y_d^{}\hat Y_d^\dagger\right)}\,.
\ee
For an individual complex Yukawa component, $P(|\hat Y_{ij}|<0.05)=0.12\%$, and
$P(|\hat Y_{ij}|>4)=0.03\%$.

\subsection{Standard Model}

\begin{figure}[t]
\centering
\includegraphics[width=.6\textwidth]{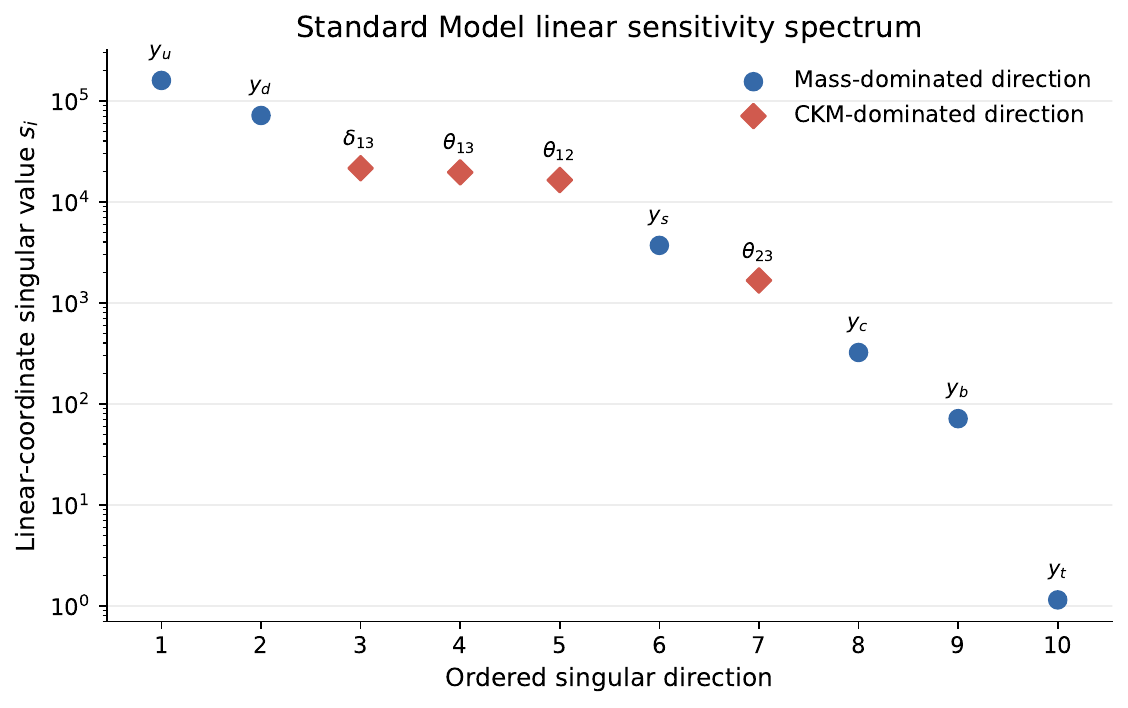}
\caption{The ten singular values of the Standard Model linear-coordinate
sensitivity matrix, ordered by size. Colors distinguish directions dominated
by quark Yukawa eigenvalues from directions dominated by CKM observables.
The mass directions are exact while the CKM directions are only approximate.}
\label{fig:SM-linear-spectrum}
\end{figure}

It is very instructive to consider the SM proper. The constancy of $h_{\rm SM}$ on $M$ (see section \ref{sec:posterior}) immediately implies constancy of $S^{}_{\rm lin}S_{\rm lin}\tra$, on $M$, and hence of all fine-tuning metrics based on its SVs. Therefore  averaging or minimization over $M$ is trivial in this case.
 Let us write the observables as \[\mu=(y_u,y_c,y_t,y_d,y_s,y_b,\t_{12},\t_{13},\t_{23},\delta_{13})\,.\]
In fact, not only is $h_{\rm SM}$ constant, it also turns out to have the simple block diagonal form
\be
h_{\rm SM}=\begin{pmatrix}
\one_{6\times 6}\\
&h_{\rm CKM}
\end{pmatrix}\,,
\ee
where the eigenvalues of the CKM block can be computed numerically.
We then find  the following singular values for $S_{\rm lin}$. From the mass block one straightforwardly gets
\be
(y_u^{-1},y_c^{-1},y_t^{-1},y_d^{-1},y_s^{-1},y_b^{-1})=(1.6\times 10^5,\ 3.2\times 10^2,\ 1.2, \ 7.1\times 10^4,\ 3.7\times 10^3,\ 71)\,,
\ee
while from the CKM block we have
\be
(2.1,\ 2.0,\ 1.6,\ 0.17)\times 10^4\,,
\ee
with left singular vectors pointing approximately in the $\delta_{13},\ \t_{13},\ \t_{12},$ and $\t_{23}$ directions respectively.
The spectrum is shown in figure \ref{fig:SM-linear-spectrum}. As expected, all but one of the singular values are very large.
Due to the constancy of the matrix $S^{}_{\rm lin}S_{\rm lin}\tra$, the average  values are exact in the SM. They are given as $s_{\rm max}=1.6\times 10^5$ for the MSV and $J=1.3\times 10^{34}$ for the PSV metric.

\begin{figure}[t]
\centering
\includegraphics[width=\textwidth]{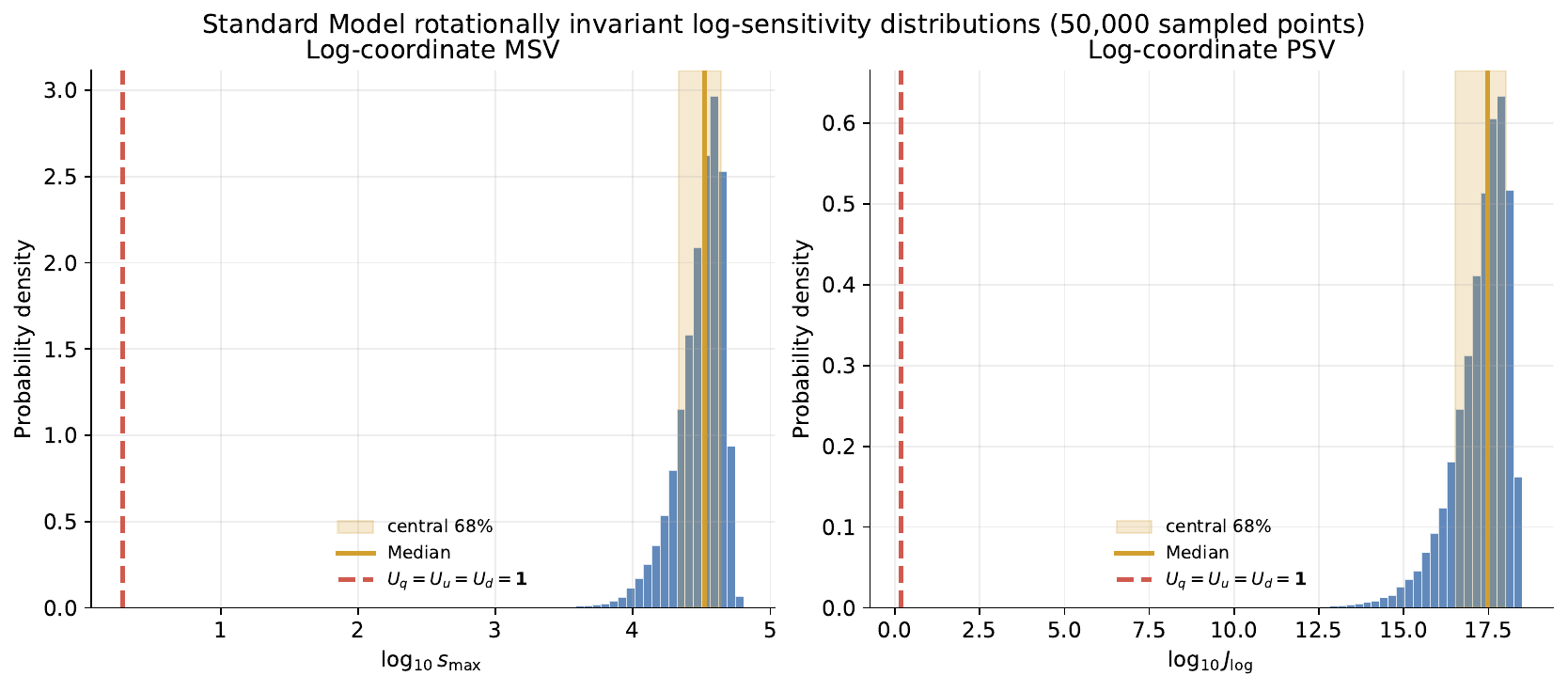}
\caption{Distributions of the logarithmic-coordinate MSV and PSV,
$J$, from $50{,}000$ points sampled on the Standard Model best-fit
manifold. The shaded bands contain the central $68\%$ of each distribution;
solid lines mark the medians. The dashed lines show the log sensitivity at $U_q=U_u=U_d=\one$.}
\label{fig:SM-log-distributions}
\end{figure}

\begin{figure}[t]
\centering
\includegraphics[width=\textwidth]{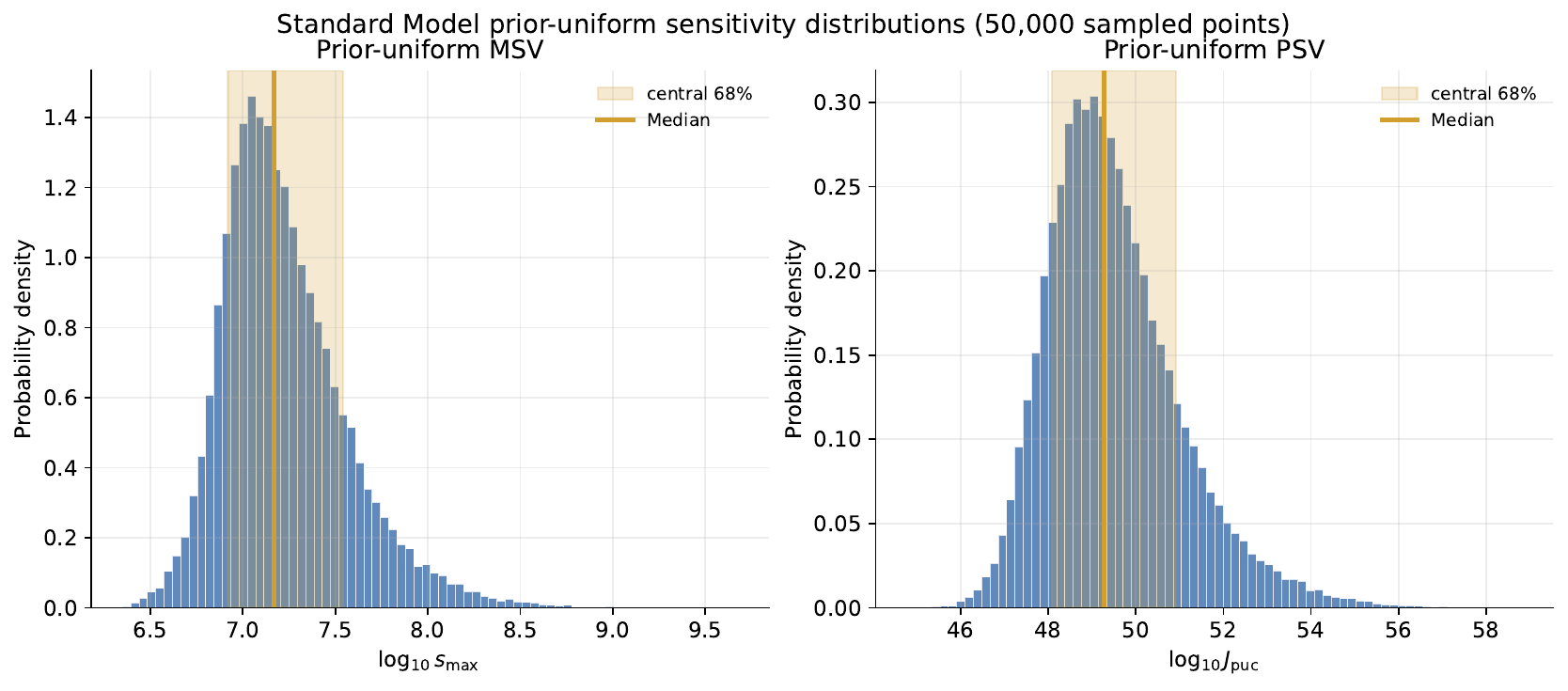}
\caption{Distributions of the prior-uniform MSV and
PSV, evaluated at the same $50{,}000$ points as in
figure~\ref{fig:SM-log-distributions}. The shaded bands contain the central
$68\%$ of each distribution and the solid lines mark the medians.
}
\label{fig:SM-puc-distributions}
\end{figure}

\begin{figure}[t]
\centering
\includegraphics[width=\textwidth]{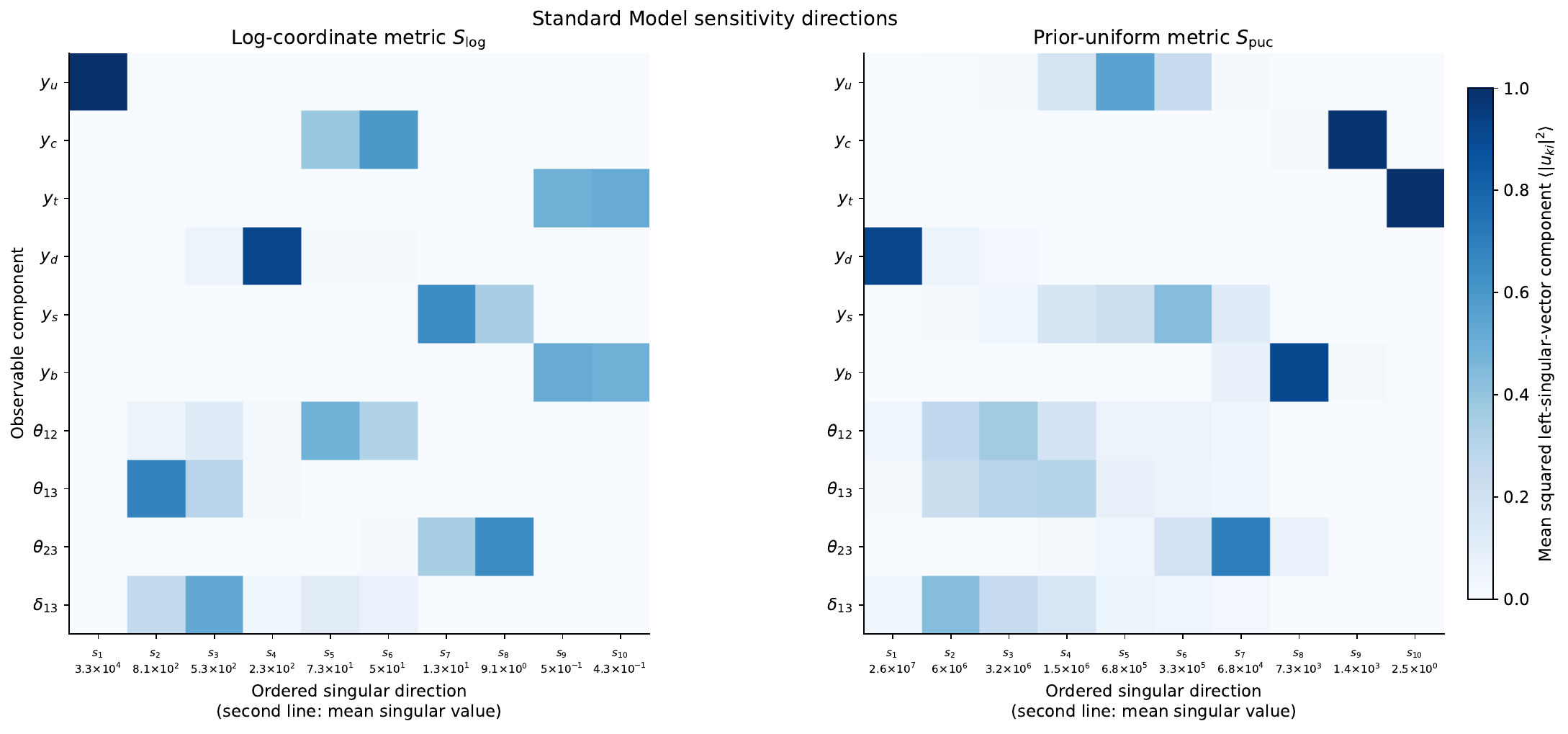}
\caption{Mean squared observable components of the left singular vectors of
$S_{\rm log}$ (left) and $S_{\rm puc}$ (right) over the sampled
Standard Model best-fit manifold. Columns are ordered by decreasing singular
value; the second line below each column gives the corresponding mean singular
value.}
\label{fig:SM-log-directions}
\end{figure}

We can also numerically evaluate the MSV and PSV metrics based on the standard logarithmic
 matrix $S_{\rm log}$, eq.~(\ref{eq:BG}). However,  $S_{\rm log}^{}S_{\rm log}\tra$ is no longer constant on $M$. Still, since the invariant measure can be sampled easily, the medians can be evaluated straightforwardly without a dedicated MCMC.
We find  $s_{\rm max}=3.3\times 10^{4}$ for the median MSV metric (aligned with the $y_u$ direction) and  $J_{\rm log}=2.9\times 10^{17}$ for the median PSV
metric.
These smaller values are expected from the exact relation $(S_{\rm log})_{ij}=(S_{\rm lin})_{ij}\theta_j$.
The typical values of $\theta_j$ are of the order of $y_t/\sqrt{18}$ for the 18 real components of $Y_u$, and similarly for $Y_d$. Hence the length of the $y_u$ row of $S_{\rm log}$ is about $y_t/\sqrt{18}\approx 0.2$ times the length of the corresponding row of $S_{\rm lin}$. This explains roughly the modest reduction  compared to the linear metrics.
There are also special points on the best-fit manifold that yield dramatic false negatives. For instance, at the special point
$U_q=U_u=U_d=\one$ we find $s_{\rm max}=1.9$ and $J=1.5$. This point grossly violates
the $O(1)$ criterion for the $Y_{ij}$.
The full MSV and PSV distributions are shown in
figures~\ref{fig:SM-log-distributions} and \ref{fig:SM-puc-distributions}, and
the average singular values and tuning directions are compared in
figure~\ref{fig:SM-log-directions}.

The prior-uniform metric gives a sharply different assessment.
For the same $50{,}000$ points we find the median values
$s_{\rm max}=1.5\times10^7$ and
$J_{\rm puc}=1.9\times10^{49}$.
The leading left singular vector has on average $91\%$ of its squared norm in
the $y_d$ direction, which is its largest component for $97\%$ of the
sampled points. The inverse radial-density factor in $S_{\rm puc}$ strongly
penalizes the small entries of the down-type Yukawa matrix.

The main SM results are summarized in table~\ref{tab:SMmetrics}.

\begin{table}[t]
\centering
\footnotesize
\setlength{\tabcolsep}{4pt}
\begin{tabular}{@{}p{1.4cm}p{2.8cm}ccp{7.1cm}@{}}
\toprule
Metric & Statistic & MSV  & PSV & Interpretation \\
\midrule
Linear
& \makecell[l]{Exact on $M$}
& $1.6\times10^5$
& $1.3\times10^{34}$
& Dominated by $y_u$.  Since $S^{}_{\rm lin}S_{\rm lin}\tra$ is
  constant, its mean, median, and minimum coincide. \\
Log
& \makecell[l]{Median on $M$}
& $3.3\times10^4$
& $2.9\times10^{17}$
& Dominated by $y_u$.  This is the prior-typical result, the smaller scores
  reflect the rescaling of the sensitivity columns by the Yukawa parameters. \\
Prior-uniform
& \makecell[l]{Median on $M$}
& $1.5\times10^7$
& $1.9\times10^{49}$
& Dominated by $y_d$. The prior-uniform coordinates penalize the small
  down-type Yukawa entries through the inverse radial prior density.
  \\
\bottomrule
\end{tabular}
\caption{Summary of the Standard Model sensitivity metrics. The first
row is independent of the point on the best-fit manifold $M$, the second and third
rows are medians over the invariant measure on $M$.
}
\label{tab:SMmetrics}
\end{table}

\clearpage
\subsection{Froggatt--Nielsen}
\label{sec:FN}
We now turn to Froggatt--Nielsen models.
In order to select a manageable number of charge assignments, we proceed as follows.
We start with all charge assignments satisfying\footnote{In principle we could consider much higher maximal charges.
However,  a model with, say, $Q_{u_i}$ charges equal to $(28,17,13)$ is arguably more difficult to UV complete than a model with $(3,2,1)$ (for instance in view of anomaly cancellation or embedding in non-Abelian groups), and hence should be penalized by introducing a prior for the charges.
Here, we simply and arbitrarily cut off our charges at $Q_{\max}=4$.}
\be
\begin{aligned}
0={}&Q_{q_3}\leq Q_{q_2}\leq Q_{q_1}\leq 4\,,\\
0\geq {}&Q_{u_3}\geq Q_{u_2}\geq Q_{u_1}\geq -4\,,\\
0\geq{}&Q_{d_3}\geq Q_{d_2}\geq Q_{d_1}\geq -4\,.
\end{aligned}
\ee
The constraint $Q_{q_3}=0$ fixes the shift redundancy $Q_{f_i}\to Q_{f_i}+const$ evident from eq.~(\ref{eq:FNcharges}).
We also retain only assignments that have greatest common divisor equal to one.
Notice that all exponents appearing in eqs.~(\ref{eq:svsFN}) and (\ref{eq:ckmFN}) are non-negative.
For a given charge assignment we estimate a
rough value  $\hat\epsilon$ by minimizing $\sum_k (\log(\mu^k_{\rm exp})-n_k\log\eps)^2$,
where $n_k$ are the exponents of $\epsilon$ appearing in the relations  $\mu^k_{\rm exp}\sim \eps^{n_k}$ as in eqs.~(\ref{eq:svsFN}) and (\ref{eq:ckmFN}), e.g.
\be
n_t=Q_{q_3}-Q_{u_3}\,,
\ee
etc.
Using this rough estimate, we determine the  parameters $\hat\mu^k\equiv\mu^k_{\rm exp}\hat \eps^{-n_k}$ and only retain charge assignments that yield $\frac{1}{5}< |\hat \mu^k|<5$.
The value of $\hat \epsilon$ is only used for this rough cut and is discarded in the following. This procedure leaves us with 165 textures.

\begin{figure}[t]
\centering
\includegraphics[width=\textwidth]{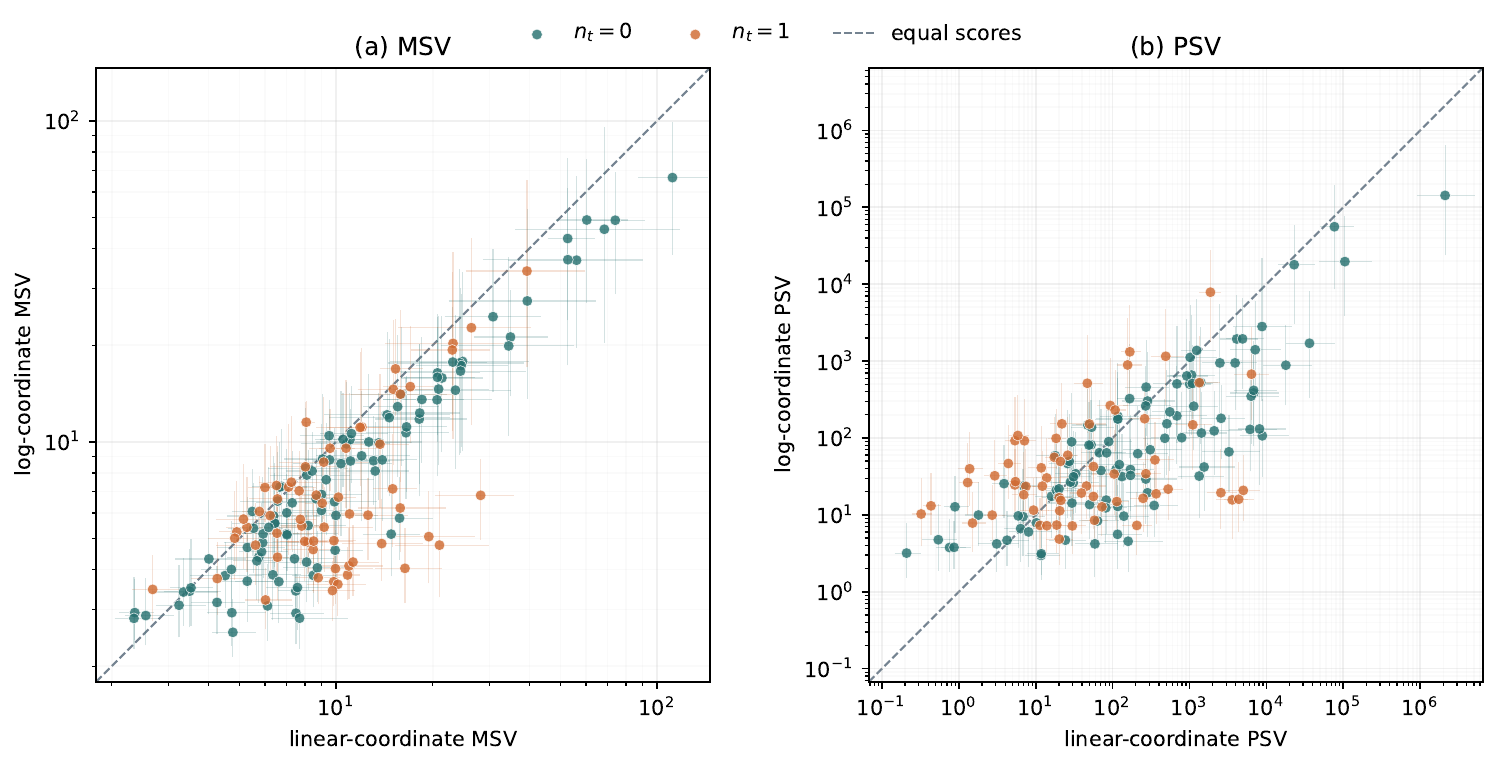}
\caption{Comparison of the linear-coordinate and logarithmic-coordinate
sensitivity metrics for the complete 165-texture scan, using the $\sigma=1$
Gaussian proto-Yukawa prior. The panels show the posterior median MSV (left)
and PSV (right). Horizontal and vertical bars show
the central 68\% posterior intervals of the linear- and log-coordinate
metrics, respectively. Colors distinguish textures with $n_t=0$ and $n_t=1$,
and the dashed lines denote equal scores.}
\label{fig:FN-linear-vs-log-metrics}
\end{figure}

We will report the MSV and PSV metrics at a fixed value of $\epsilon$ (that we will refer to as $\eps_\star$)  determined as follows. For a given charge assignment $Q$, we consider the evidence of the experimental data given $\epsilon$,
\be
p_Q(\mu_{\rm exp}|\eps)
	\sim \eps^{-Q_{\rm tot}}  \int_M e^{-V_{Q,\eps}(u)} dH(u)\,,
\ee
where the constant of proportionality is independent of both $Q$ and $\epsilon$, and is therefore common to all charge assignments compared below.
For any given charge assignment we then numerically maximize $p_Q(\mu_{\rm exp}|\eps)$ to find the optimal $\eps$.\footnote{We have checked in select cases that this value is virtually identical with the median of the posterior distribution
$\sim p_Q(\mu_{\rm exp}|\eps)p(\eps) $ using a uniform prior with $0\leq \eps\leq 1$.}
Throughout the scan we use the unit-width Gaussian prior for the real and
imaginary parts of the proto-Yukawa couplings. Figure~\ref{fig:FN-linear-vs-log-metrics}
compares the linear-coordinate and logarithmic-coordinate sensitivity metrics.
We will work with this optimal $\eps_\star$ parameter in each model. However, for the BF we  report a marginalized evidence, using a uniform prior with $0\leq \eps\leq 1$.

The definition of the matrix $S_\text{log}$ is slightly ambiguous, as we might differentiate with respect to the logarithm of the real and imaginary parts of  $\hat Y$, or with respect to the logarithm of the complex variable $\hat Y$. The latter choice, which we adopt throughout this paper, is equivalent to differentiating with respect to $\log |\hat Y_{ij}|$ and $\arg \hat Y_{ij}$, see appendix \ref{sec:complex}.

For FN models, the matrix $S^{}_{\rm lin}S_{\rm lin}\tra$ is no longer constant along the best-fit manifold. Therefore, both the linear- and log-coordinate sensitivity metrics
acquire nontrivial distributions on the best-fit manifold. Figure~\ref{fig:FN-linear-vs-log-metrics}
shows that the linear and logarithmic prescriptions remain correlated, but do
not induce identical rankings. The agreement is stronger for the MSV
($\rho=0.77$) than for the PSV ($\rho=0.66$), using Spearman rank
correlations (defined as the correlation of the rankings derived from each measure)  of the 165 textures.

\begin{figure}[t]
\centering
\includegraphics[width=\textwidth]{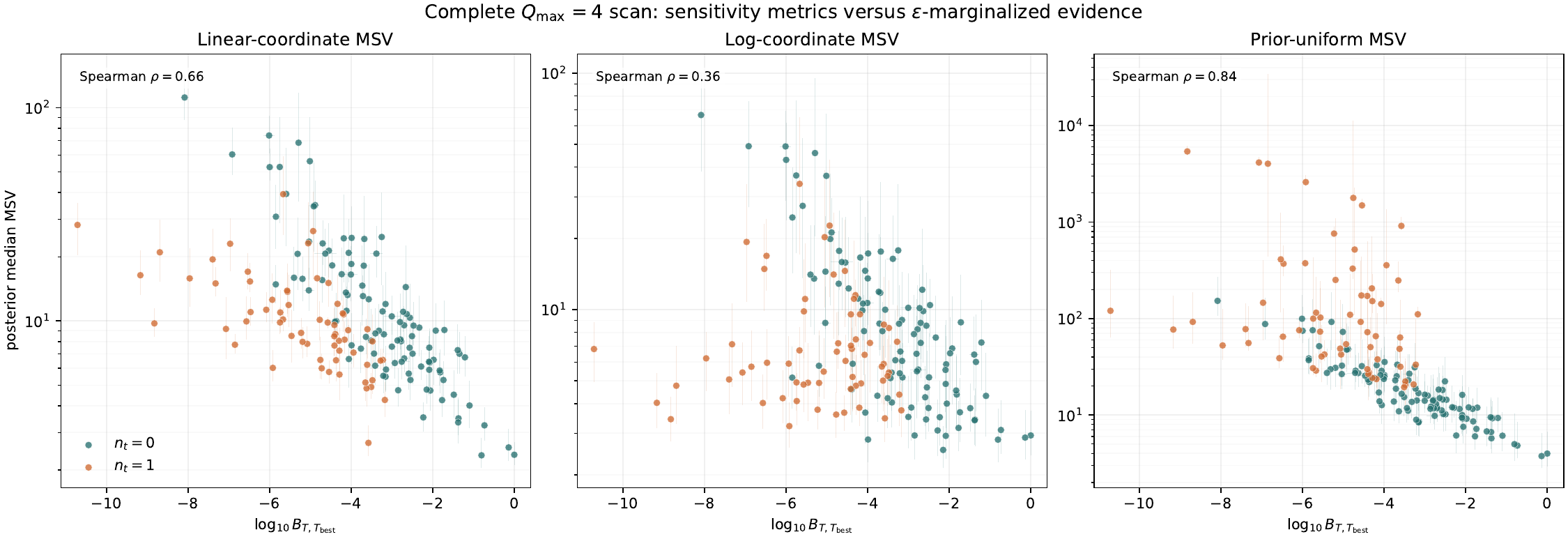}
\caption{Posterior median linear-coordinate (left), logarithmic-coordinate
(center), and prior-uniform (right) MSVs versus the
$\epsilon$-marginalized Bayes factor for the complete 165-texture
$Q_{\max}=4$ scan.  The evidence is normalized to the best texture, so that
$\log_{10}B_{T,T_{\rm best}}=0$ for that texture.  Vertical bars show the
central 68\% posterior intervals of the MSV distributions.  Colors distinguish textures with
$n_t=0$ and $n_t=1$.}
\label{fig:FN-MSV-vs-BF}
\end{figure}

The complete scan of 165 textures allows us to compare the local sensitivity
metrics directly with the Bayesian evidence.  Figure~\ref{fig:FN-MSV-vs-BF}
shows the posterior median linear-coordinate, logarithmic-coordinate, and
prior-uniform MSVs as functions of the Bayes factor relative to
the best texture in the scan.  All three metrics are correlated with the
evidence, as expected if they capture a common notion of naturalness.
The Spearman correlation coefficients for the linear,
logarithmic, and prior-uniform metrics are $0.66$, $0.36$, and $0.84$,
respectively.  Several textures have
moderate logarithmic MSV but very small BF. Their linear MSV is larger
and therefore more consistent with the evidence ranking.
Assuming that the prior is chosen adequately (so the BF does not produce false positives),
 the log MSV (and, to a lesser extent, the linear MSV) metric produces a good amount of  false negatives. For instance, in the 165 model scan, there are 101 models with log MSV medians  smaller than 10 but BF smaller than 0.01.
For the linear measure the corresponding number is lowered to 76, still uncomfortably high for a reliable diagnostic of naturalness.
When applying the same diagnostic to the prior-uniform metric, this number drops to 9, indicating that it produces notably fewer false negatives.
Conversely, there are zero models in this scan with BF greater than 0.01 and either linear or log MSVs greater than 10. This points to a low false negative rate for the BF metric, and a low false positive rate for the MSV metrics.
Applying the same criterion to the prior-uniform MSV, we find 3 models, each of which are rather close to both thresholds.

The $n_t=1$ and $n_t=0$ subsets show a very distinct behavior. None of the top 20 models ranked by evidence, prior-uniform coordinate MSV or  PSV (prior sensitive metrics)
have $n_t=1$. However, when ranked by the log or linear PSV or MSV (prior insensitive metrics) some models with $n_t=1$ do make it to the top 20.
The reason is that since $y_t\approx |(\hat Y_u)_{33}| \eps^{n_t}$,  $n_t=1$ requires $|(\hat Y_u)_{33}|$ of order 5 to 10 (depending on $\eps$), which is strongly suppressed by the prior.
However, from the approximate expression for $y_t$ it is evident that no anomalous sensitivity to proto Yukawas is associated to this. We can then interpret this  either as a false negative of the
prior insensitive metrics (failure to identify prior-unlikely parameter points), or as a false positive for the prior sensitive metrics (for instance, if one would like to consider say
$|(\hat Y_u)_{33}|\sim 8$ still somewhat natural, the unit Gaussian prior would be inadequate).

\begin{figure}[t]
\centering
\includegraphics[width=\textwidth]{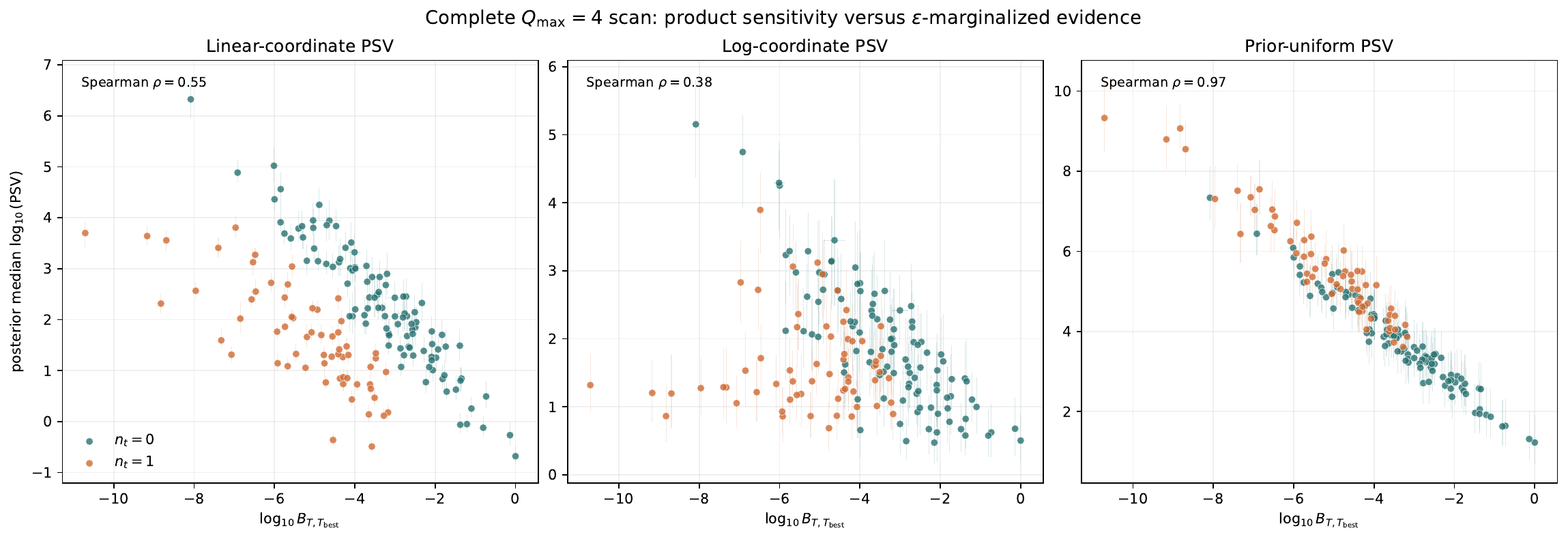}
\caption{Posterior median linear-coordinate (left), logarithmic-coordinate
(center), and prior-uniform (right) $\log_{10}(\mathrm{PSV})$
versus the $\epsilon$-marginalized Bayes factor for the complete 165-texture
 scan.  The evidence is normalized to the best texture.
Vertical bars show the central 68\% posterior intervals of the PSV
distributions.  Colors
distinguish textures with $n_t=0$ and $n_t=1$.  }
\label{fig:FN-PSV-vs-BF}
\end{figure}

\begin{figure}[t]
\centering
\includegraphics[width=0.52\textwidth]{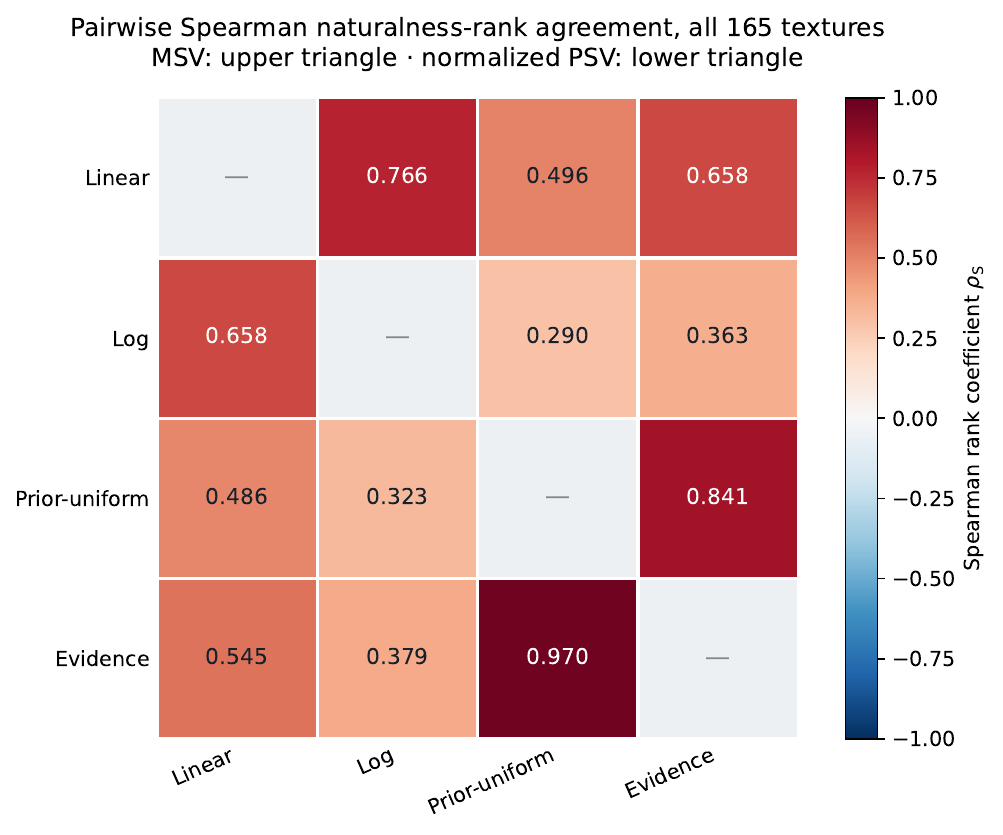}
\caption{Pairwise Spearman rank correlations among the linear-coordinate,
logarithmic-coordinate, prior-uniform, and evidence metrics for
the 165 textures.  The upper triangle contains the MSV correlations and the
lower triangle the  PSV correlations.}
\label{fig:FN-metric-correlations}
\end{figure}
\FloatBarrier

The linear and logarithmic PSV rankings have Spearman correlations $0.55$ and
$0.38$ with the evidence, while the prior-uniform PSV gives
$\rho=0.97$.  The correlation matrix in
figure~\ref{fig:FN-metric-correlations} shows that the prior-uniform metric
agrees most strongly with the evidence for both MSV and PSV.
The reason for this can be traced back to the observation made after eq.~(\ref{eq:observation}).

The possible suppression of the PSV by anomalously small singular values is not very common
 in this scan.  SVs smaller than one do occur, but they are not
parametrically small.  Across the pooled posterior samples, the median values
of the smallest singular value are $0.30$, $0.47$, and $0.58$ for the linear,
logarithmic, and prior-uniform coordinates, respectively.
Replacing every
$s_a<1$ by one changes the normalization but preserves the texture ranking: the Spearman coefficients between the ordinary and clipped PSV rankings
are $0.955$, $0.995$, and $0.998$ in the linear, logarithmic, and prior-uniform
cases.  Since no very small singular direction is present, we use the
unmodified PSV and do not apply a clipping prescription.

Table~\ref{tab:FN-top20} collects the numerical results for the 20 textures
with the largest evidence and the 10 with the smallest evidence after
marginalizing over $\epsilon$.
We give, for comparison, the ranking of ref.~\cite{Cornella:2023zme}.
That work characterizes a texture by the fraction of randomly drawn order-one
coefficients for which all quark masses and mixing observables are reproduced
within a factor of two or five.  These success fractions test whether a texture
can reproduce the observed hierarchies over an appreciable part of its prior
parameter space, but a broad acceptance window need not measure accurately how
much probability is concentrated near the measured  point.
In fact, some textures (such as their 9, 16, and 19) rank
 appreciably lower in our evidence analysis.
In such cases, we checked (using the same priors as in \cite{Cornella:2023zme})  that  draws  passing the factor-2 criterion tend to accumulate near
the boundary of the window and commonly miss several observables by
order-one factors at the same time.  When the allowed mismatch is reduced from
a factor of two to a factor of $1.5$, their accepted fractions decrease more
rapidly than for the highest-ranked texture.
Correlations among the predicted observables can therefore make an extended
region of parameter space pass the factor-2 cut without bringing it particularly
close to the measured point. \footnote{Optimizing $\epsilon$ separately for every draw, as done in \cite{Cornella:2023zme},
also increases the success fractions, because this procedure does
not penalize a texture for requiring a restricted range of $\epsilon$, although
this is not the dominant effect in the examples studied.}  Finally, the
comparison depends on the coefficient prior. Ref.~\cite{Cornella:2023zme} uses
log-normal distributions for the magnitudes, whereas our main analysis uses normal distributions  for the real
and imaginary parts.  Changing between these priors can shift relative
evidences by approximately one order of magnitude.  Thus the ranking can change
substantially, even though the best textures remain stable under many of the
criteria considered.
Regarding ref.~\cite{Ibe:2024cvi}, a direct comparison is complicated by the fact that the authors used GUT scale values for the Yukawa couplings, which tend to be reduced by a factor of two due to
the universal QCD contribution. However, we did run some of our textures with the GUT scale data used in \cite{Ibe:2024cvi}, finding perfect agreement for the Bayes factors in all cases.

\begin{table}[p]
\centering
\scriptsize
\setlength{\tabcolsep}{2.5pt}
\begin{tabular}{@{}cc ccc cc ccc @{\hspace{4pt}} ccc@{}}
\toprule
\multicolumn{2}{c}{Rank}&\multicolumn{4}{c}{Texture}&& \multicolumn{3}{c}{$\log_{10}(\mathrm{MSV})$}
& \multicolumn{3}{c}{$\log_{10}(\mathrm{PSV})$} \\
\cmidrule(lr){1-2}\cmidrule(lr){3-6}\cmidrule(lr){8-10}\cmidrule(lr){11-13}
by BF & \cite{Cornella:2023zme} & $Q_q$ & $\lvert Q_u\rvert$ & $\lvert Q_d\rvert$ & $\epsilon_\star$
& $\log_{10}(\text{BF})$
& lin & log & puc & lin & log & puc \\
\midrule
1 & 1 & $(3,2,0)$ & $(4,2,0)$ & $(3,3,3)$ & 0.168 & 0.00 & 0.37 & 0.47 & 0.60 & $-0.68$ & 0.50 & 1.23 \\
2 & 2 & $(3,2,0)$ & $(4,2,0)$ & $(4,3,3)$ & 0.178 & $-0.14$ & 0.41 & 0.46 & 0.57 & $-0.27$ & 0.68 & 1.31 \\
3 & 3 & $(3,2,0)$ & $(3,1,0)$ & $(3,2,2)$ & 0.120 & $-0.73$ & 0.51 & 0.49 & 0.68 & 0.49 & 0.62 & 1.64 \\
4 & 5 & $(4,3,0)$ & $(4,2,0)$ & $(4,3,3)$ & 0.228 & $-0.80$ & 0.37 & 0.45 & 0.70 & $-0.12$ & 0.58 & 1.63 \\
5 & 13 & $(4,3,0)$ & $(4,2,0)$ & $(4,4,3)$ & 0.244 & $-1.10$ & 0.60 & 0.63 & 0.79 & 0.25 & 1.00 & 1.87 \\
6 & -- & $(3,2,0)$ & $(3,2,0)$ & $(3,3,3)$ & 0.167 & $-1.21$ & 0.83 & 0.86 & 0.97 & $-0.05$ & 1.10 & 1.92 \\
7 & 4 & $(3,2,0)$ & $(4,1,0)$ & $(3,3,3)$ & 0.163 & $-1.34$ & 0.85 & 0.78 & 0.97 & 0.84 & 1.37 & 2.56 \\
8 & 7 & $(4,2,0)$ & $(4,2,0)$ & $(4,3,3)$ & 0.209 & $-1.37$ & 0.54 & 0.53 & 0.76 & 0.80 & 0.82 & 2.06 \\
9 & 15 & $(4,3,0)$ & $(4,2,0)$ & $(3,3,3)$ & 0.222 & $-1.38$ & 0.52 & 0.53 & 0.82 & $-0.06$ & 0.58 & 1.94 \\
10 & 6 & $(3,2,0)$ & $(4,1,0)$ & $(3,3,2)$ & 0.145 & $-1.39$ & 0.86 & 0.81 & 0.98 & 1.49 & 1.42 & 2.58 \\
11 & 8 & $(3,2,0)$ & $(3,1,0)$ & $(2,2,2)$ & 0.114 & $-1.48$ & 0.65 & 0.58 & 0.83 & 0.63 & 0.67 & 1.97 \\
12 & -- & $(4,3,0)$ & $(4,2,0)$ & $(4,4,4)$ & 0.266 & $-1.71$ & 0.96 & 0.95 & 1.08 & 0.59 & 1.41 & 2.44 \\
13 & 14 & $(3,2,0)$ & $(4,2,0)$ & $(4,3,2)$ & 0.167 & $-1.74$ & 0.72 & 0.57 & 0.86 & 1.47 & 1.15 & 2.70 \\
14 & 10 & $(3,2,0)$ & $(4,2,0)$ & $(3,3,2)$ & 0.158 & $-1.77$ & 0.63 & 0.50 & 0.78 & 0.91 & 0.78 & 2.52 \\
15 & 11 & $(2,1,0)$ & $(3,1,0)$ & $(2,2,2)$ & 0.063 & $-1.82$ & 0.76 & 0.64 & 0.93 & 0.83 & 0.98 & 2.60 \\
16 & -- & $(4,3,0)$ & $(4,1,0)$ & $(4,3,2)$ & 0.206 & $-1.83$ & 0.77 & 0.66 & 1.06 & 1.70 & 1.13 & 2.82 \\
17 & -- & $(3,2,0)$ & $(4,1,0)$ & $(4,3,3)$ & 0.176 & $-1.93$ & 0.95 & 0.84 & 1.08 & 1.41 & 1.67 & 2.89 \\
18 & -- & $(3,2,0)$ & $(4,2,0)$ & $(4,4,3)$ & 0.203 & $-1.98$ & 0.82 & 0.82 & 0.96 & 1.25 & 1.77 & 2.73 \\
19 & 18 & $(4,2,0)$ & $(4,2,0)$ & $(3,3,3)$ & 0.204 & $-2.06$ & 0.67 & 0.60 & 0.88 & 1.01 & 0.90 & 2.37 \\
20 & 9 & $(3,2,0)$ & $(3,1,0)$ & $(3,3,2)$ & 0.140 & $-2.07$ & 0.81 & 0.77 & 0.96 & 1.52 & 1.53 & 2.78 \\
\cmidrule(lr){1-13}
156 & -- & $(4,2,0)$ & $(3,1,1)$ & $(4,4,2)$ & 0.215 & $-6.97$ & 1.36 & 1.29 & 2.16 & 3.81 & 2.83 & 7.04 \\
157 & -- & $(3,2,0)$ & $(4,2,1)$ & $(3,3,2)$ & 0.178 & $-7.07$ & 0.96 & 0.73 & 3.62 & 1.31 & 1.05 & 7.35 \\
158 & -- & $(4,2,0)$ & $(4,3,1)$ & $(3,3,3)$ & 0.237 & $-7.32$ & 1.18 & 0.85 & 1.75 & 1.59 & 1.28 & 6.43 \\
159 & -- & $(4,2,0)$ & $(4,3,1)$ & $(4,4,2)$ & 0.250 & $-7.40$ & 1.29 & 0.70 & 1.89 & 3.41 & 1.29 & 7.51 \\
160 & -- & $(4,3,0)$ & $(4,2,1)$ & $(3,3,2)$ & 0.238 & $-7.96$ & 1.20 & 0.79 & 1.72 & 2.57 & 1.27 & 7.31 \\
161 & -- & $(2,2,0)$ & $(3,0,0)$ & $(4,2,2)$ & 0.120 & $-8.08$ & 2.05 & 1.82 & 2.18 & 6.32 & 5.15 & 7.34 \\
162 & -- & $(4,2,0)$ & $(4,3,1)$ & $(4,3,2)$ & 0.239 & $-8.69$ & 1.32 & 0.68 & 1.97 & 3.56 & 1.19 & 8.55 \\
163 & -- & $(3,2,0)$ & $(4,2,1)$ & $(4,2,2)$ & 0.176 & $-8.82$ & 0.99 & 0.54 & 3.73 & 2.31 & 0.86 & 9.07 \\
164 & -- & $(4,3,0)$ & $(4,2,1)$ & $(4,2,2)$ & 0.238 & $-9.17$ & 1.21 & 0.61 & 1.89 & 3.64 & 1.20 & 8.80 \\
165 & -- & $(4,3,0)$ & $(4,2,1)$ & $(3,2,2)$ & 0.235 & $-10.71$ & 1.45 & 0.83 & 2.08 & 3.70 & 1.32 & 9.33 \\
\cmidrule(lr){1-13}
&&\multicolumn{4}{c}{SM} & $-84.77$ & 5.20 & 4.52 & 7.18 & 34.11 & 17.46 & 49.28 \\
\bottomrule
\end{tabular}
\caption{The 20 highest-evidence and 10 lowest-evidence textures in the
complete $Q_{\max}=4$ scan, ranked by the $\epsilon$-marginalized evidence.
The evidence is shown as
$\log_{10}B_{T,T_{\rm best}}$ relative to the best texture (the first row).  The
MSV and  PSV entries are posterior medians
in linear, logarithmic, and prior-uniform coordinates at fixed $\eps_\star$.  The final row gives
the corresponding Standard Model values for comparison.  The second column quotes the ranks
of the same textures according to ref.~\cite{Cornella:2023zme}.}
\label{tab:FN-top20}
\end{table}

Amongst our 165 quark textures there are  ten assignments compatible with the quark-sector SU$(5)$ condition
$Q_q=\lvert Q_u\rvert$, they are collected separately in table~\ref{tab:FN-SU5}.   These data should be taken only as a rough baseline for a more detailed GUT analysis.
The latter should take into account the fact that the lepton charges are fixed in terms of the quark charges. Moreover, a strong correlation between $Y_d$ and $Y_e$ is expected
from the exact GUT relation $Y^{}_e=Y_d\tra$, up to some threshold corrections and potentially additional GUT breaking effects.
Finally, the effects of running the Yukawa couplings from the TeV to the GUT scale, potentially incorporating superpartner contributions is another non-negligible effect.
We leave a dedicated GUT analysis to future work.

\begin{table}[p]
\centering
\scriptsize
\setlength{\tabcolsep}{2.5pt}
\begin{tabular}{@{}cc ccc cc ccc @{\hspace{4pt}} ccc@{}}
\toprule
\multicolumn{2}{c}{Rank}&\multicolumn{4}{c}{Texture}&& \multicolumn{3}{c}{$\log_{10}(\mathrm{MSV})$}
& \multicolumn{3}{c}{$\log_{10}(\mathrm{PSV})$} \\
\cmidrule(lr){1-2}\cmidrule(lr){3-6}\cmidrule(lr){8-10}\cmidrule(lr){11-13}
by BF & \cite{Cornella:2023zme} & $Q_q$ & $\lvert Q_u\rvert$ & $\lvert Q_d\rvert$ & $\epsilon_\star$
& $\log_{10}B$
& lin & log & puc & lin & log & puc \\
\midrule
6 & -- & $(3,2,0)$ & $(3,2,0)$ & $(3,3,3)$ & 0.167 & $-1.21$ & 0.83 & 0.86 & 0.97 & $-0.05$ & 1.10 & 1.92 \\
8 & 7 & $(4,2,0)$ & $(4,2,0)$ & $(4,3,3)$ & 0.209 & $-1.37$ & 0.54 & 0.53 & 0.76 & 0.80 & 0.82 & 2.06 \\
19 & 18 & $(4,2,0)$ & $(4,2,0)$ & $(3,3,3)$ & 0.204 & $-2.06$ & 0.67 & 0.60 & 0.88 & 1.01 & 0.90 & 2.37 \\
21 & -- & $(4,2,0)$ & $(4,2,0)$ & $(4,3,2)$ & 0.191 & $-2.08$ & 0.87 & 0.47 & 1.21 & 1.77 & 0.62 & 2.92 \\
23 & 17 & $(4,2,0)$ & $(4,2,0)$ & $(4,4,3)$ & 0.226 & $-2.09$ & 0.81 & 0.74 & 0.97 & 1.27 & 1.32 & 2.57 \\
25 & 20 & $(3,2,0)$ & $(3,2,0)$ & $(3,3,2)$ & 0.149 & $-2.23$ & 0.55 & 0.54 & 0.79 & 0.77 & 0.99 & 2.65 \\
47 & -- & $(4,2,0)$ & $(4,2,0)$ & $(3,3,2)$ & 0.187 & $-3.00$ & 0.88 & 0.55 & 1.21 & 2.06 & 0.75 & 3.34 \\
51 & -- & $(2,1,0)$ & $(2,1,0)$ & $(2,2,2)$ & 0.061 & $-3.16$ & 1.08 & 0.96 & 1.23 & 1.49 & 1.50 & 3.43 \\
79 & -- & $(3,2,0)$ & $(3,2,0)$ & $(4,2,2)$ & 0.145 & $-3.76$ & 0.87 & 0.64 & 1.14 & 2.09 & 1.65 & 3.87 \\
103 & -- & $(2,1,0)$ & $(2,1,0)$ & $(2,2,1)$ & 0.046 & $-4.39$ & 1.00 & 0.66 & 1.37 & 3.13 & 1.50 & 4.53 \\
\bottomrule
\end{tabular}
\caption{The ten quark-sector SU$(5)$-compatible textures, satisfying
$Q_q=\lvert Q_u\rvert$, ordered by their $\epsilon$-marginalized evidence.
The
conventions are as in table~\ref{tab:FN-top20}.}
\label{tab:FN-SU5}
\end{table}

\subsection{Characterization of the most natural FN textures}

Inspecting the  texture ranking in table \ref{tab:FN-top20}  it is interesting to note that already the
fifth best texture loses an order of magnitude in evidence with respect to the best (a Bayes factor of 1/10 is usually
considered to  indicate a substantially disfavored model). Judging by evidence, we can therefore rather sharply characterize
the most natural textures.

\begin{figure}[t]
\centering
\includegraphics[width=0.55\textwidth]{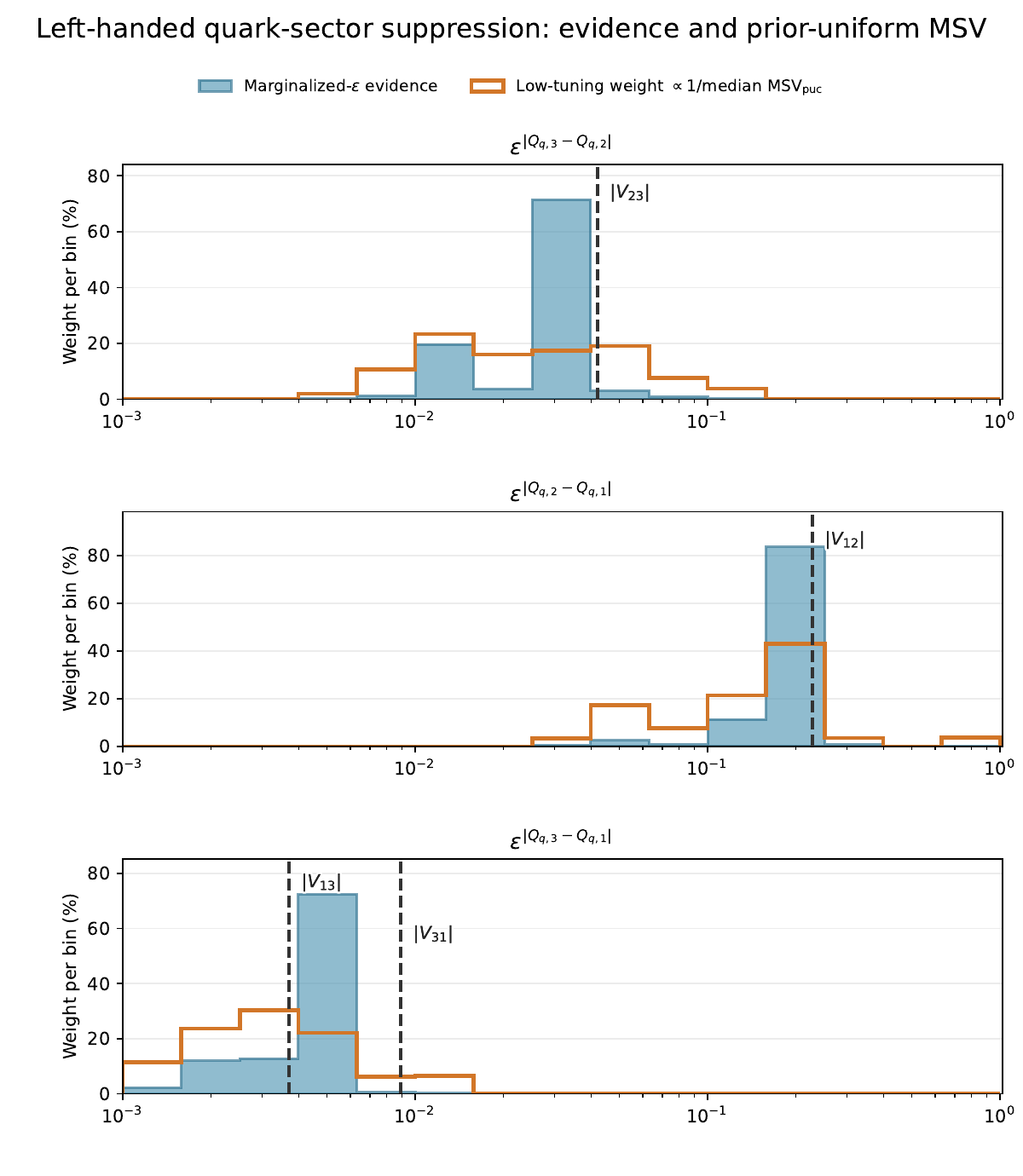}
\caption{Distributions of the left-handed quark-sector suppression factors
$\eps^{|Q_{q_3}-Q_{q_2}|}$, $\eps^{|Q_{q_2}-Q_{q_1}|}$, and
$\eps^{|Q_{q_3}-Q_{q_1}|}$ (from top to bottom) for the full 165-texture scan,
using the fitted value of $\eps$ for each texture.  Filled histograms weight
each texture by its marginalized-$\eps$ Bayesian evidence, while the outlines
use the low-tuning weight
$1/\operatorname{median}(\mathrm{MSV}_{\rm puc})$.}
\label{fig:FN-q-suppression}
\end{figure}

\begin{figure}[t]
\centering
\includegraphics[width=0.55\textwidth]{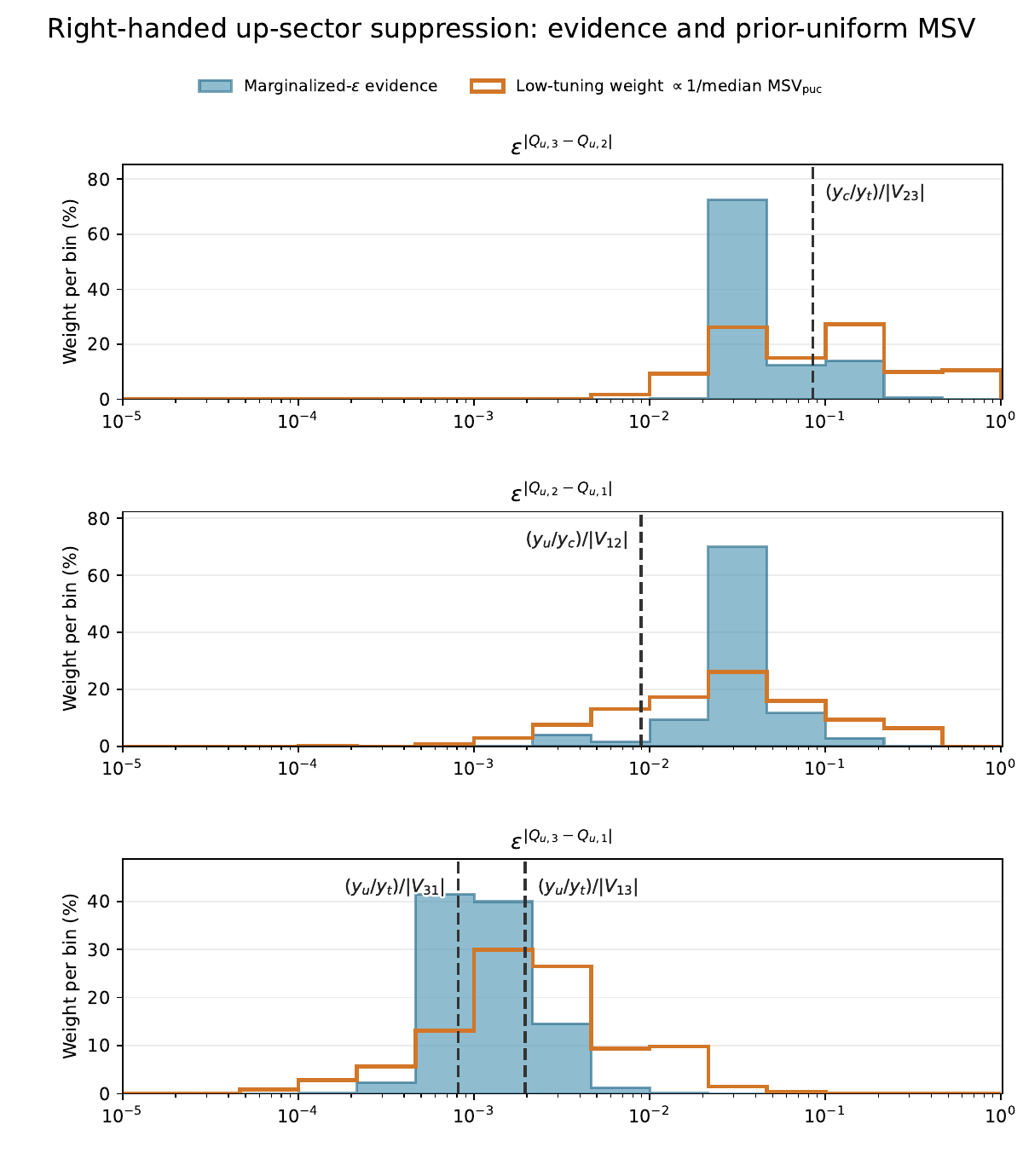}
\caption{Distributions of the right-handed up-sector suppression factors
$\eps^{|Q_{u_3}-Q_{u_2}|}$, $\eps^{|Q_{u_2}-Q_{u_1}|}$, and
$\eps^{|Q_{u_3}-Q_{u_1}|}$ (from top to bottom).  The scan and weighting
conventions are the same as in figure~\ref{fig:FN-q-suppression}.}
\label{fig:FN-u-suppression}
\end{figure}

\begin{figure}[t]
\centering
\includegraphics[width=0.55\textwidth]{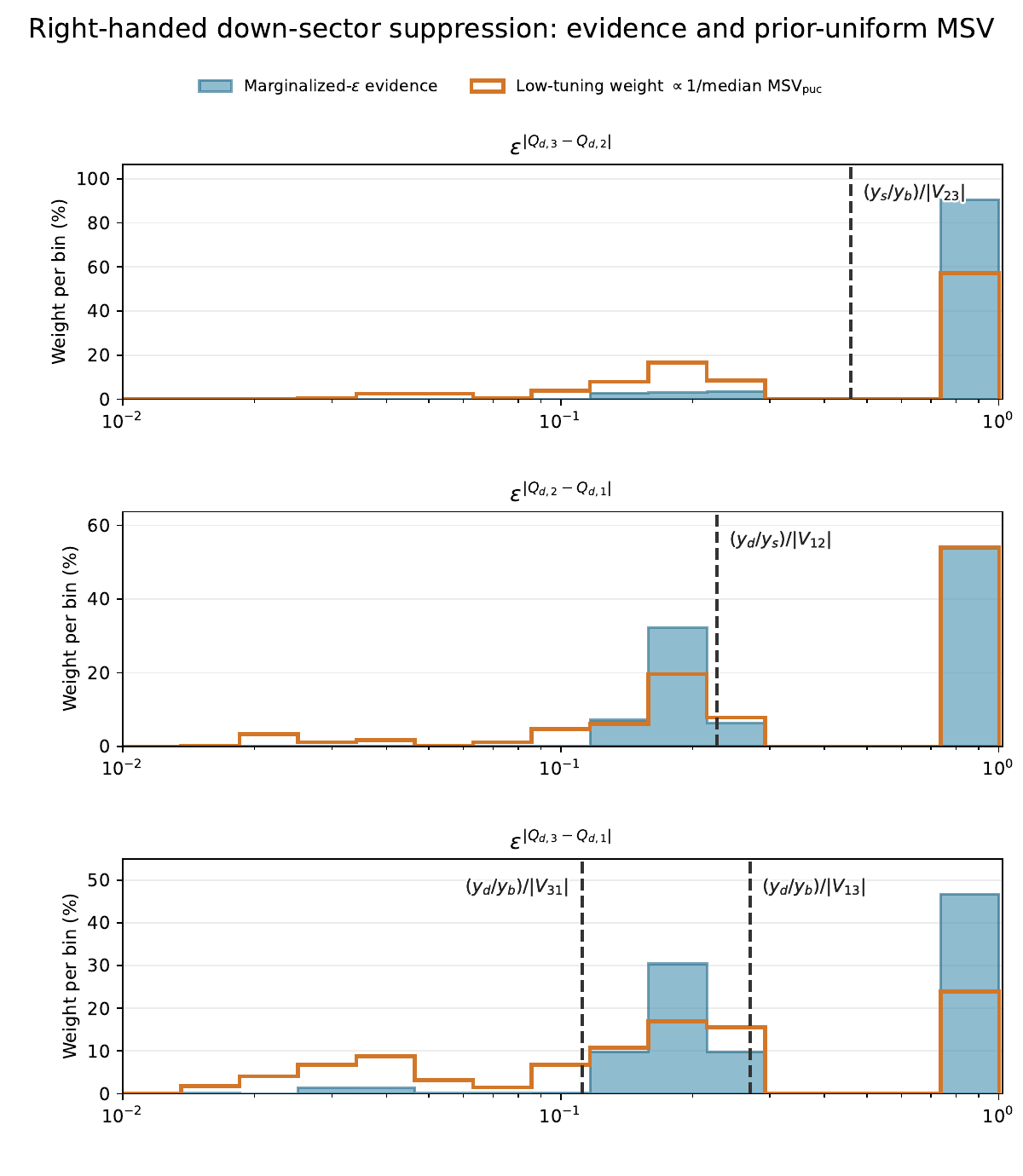}
\caption{Distributions of the right-handed down-sector suppression factors
$\eps^{|Q_{d_3}-Q_{d_2}|}$, $\eps^{|Q_{d_2}-Q_{d_1}|}$, and
$\eps^{|Q_{d_3}-Q_{d_1}|}$ (from top to bottom).  The scan and weighting
conventions are the same as in figure~\ref{fig:FN-q-suppression}.  Both measures
favor mild right-handed down-quark hierarchies.}
\label{fig:FN-down-suppression}
\end{figure}

Figures~\ref{fig:FN-q-suppression}--\ref{fig:FN-down-suppression} show the
distributions of the suppression factors
$\eps^{|Q_{f_i}-Q_{f_j}|}$, for $f=q,u,d$, over the full set of 165 textures, weighted by both evidence and prior-uniform MSV.
The evidence-weighted left-handed suppressions are clearly concentrated near the
familiar CKM angles, eq.~(\ref{eq:ckmFN}).
In contrast, the  right-handed patterns deviate somewhat from the expected FN scaling.
Moreover, the right-handed down-quark charge
hierarchy is strongly preferred to be mild.  Weighting the
textures by their Bayesian evidence gives a strong preference for
$Q_{d_2}=Q_{d_3}$.  Weighting them by the inverse median prior-uniform MSV leads
to the same qualitative conclusion, although less strongly.
As already remarked above, the value $n_t=0$ ($Q_{u_3}=Q_{q_3}=0$) is strongly preferred by all metrics. The most natural
texture (by evidence) with $n_t=1$ has a BF of $\sim 10^{-3.2}$ with respect to the best texture. This conclusion would change if we considered the lower
GUT-scale top Yukawa coupling.

\subsection{Comparing FN to the SM}

\begin{figure}[t]
\centering
\includegraphics[width=\textwidth]{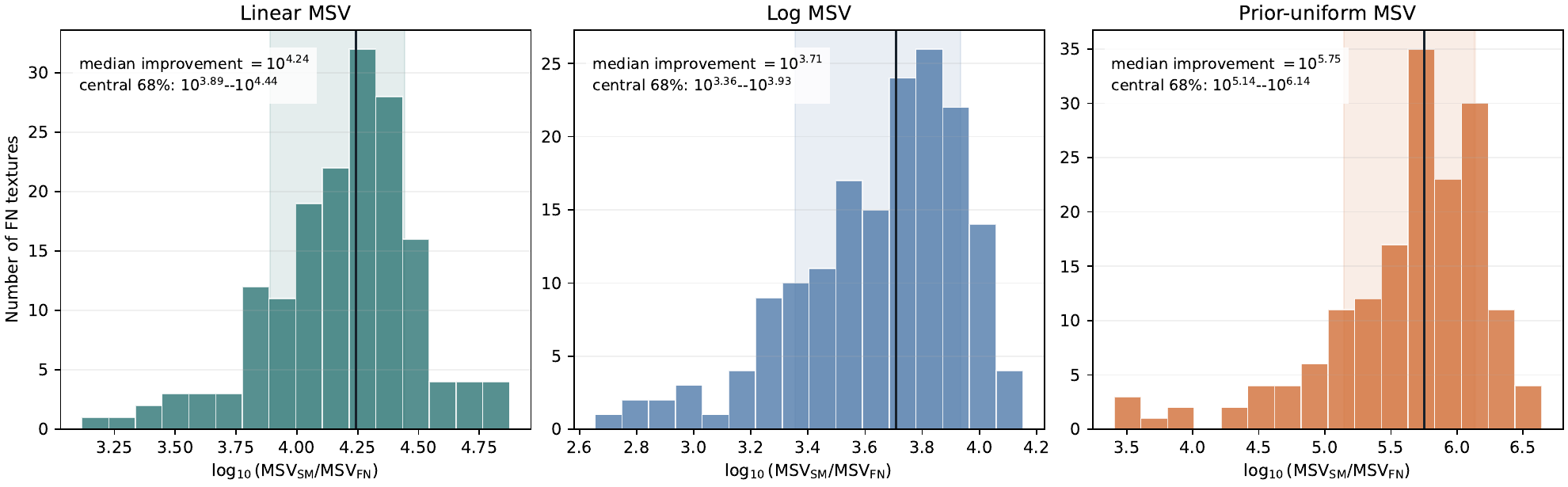}
\caption{Distributions over the 165 FN textures of the MSV improvement factor $R_X$ relative to the SM, for the linear (left),
logarithmic (center), and prior-uniform (right) parameter-space metrics.  The
black vertical lines mark the medians and the shaded regions show the central
68\% intervals.  Every texture has $R_X>1$ in all three metrics.}
\label{fig:SM-vs-FN-MSV-improvement}
\end{figure}

The sensitivity measures give a strong separation between the FN
textures and the SM.  Using the same observable normalization and the same
parameter-space metric in each comparison, all 165 FN textures have smaller
posterior median MSV than the SM values in table~\ref{tab:SMmetrics}.  For a
metric $X$, define the improvement factor of texture $T$ by
\be
R_X(T)=\frac{\operatorname{MSV}^{\rm SM}_X}
{\operatorname{MSV}^{T}_X}\,.
\ee
The median improvement factors across the complete texture scan are
\be
\operatorname{median}_T R_X=
\begin{cases}
10^{4.24}, & X={\rm linear},\\
10^{3.71}, & X={\rm logarithmic},\\
10^{5.75}, & X={\rm prior\mathchar`-uniform}\,.
\end{cases}
\ee
The distributions of these improvement factors are shown in
figure~\ref{fig:SM-vs-FN-MSV-improvement}.
Thus the conclusion that FN is substantially less sensitive than the SM is
independent of which of the three parameter-space metrics is used.

The evidence provides a complementary, explicitly prior-dependent comparison,
as discussed in section~\ref{sec:naturalness-metrics}.  For the highest-evidence
texture T1, marginalizing over $\eps$ with the normalized uniform prior on
$0<\eps<1$ gives
\be
\log_{10}B_{{\rm T1},{\rm SM}}=84.77\,.
\ee
This is the Bayes factor in eq.~(\ref{eq:FN-SM-BF}) for T1 regarded as a fixed
model.

\begin{figure}[t]
\centering
\includegraphics[width=0.62\textwidth]{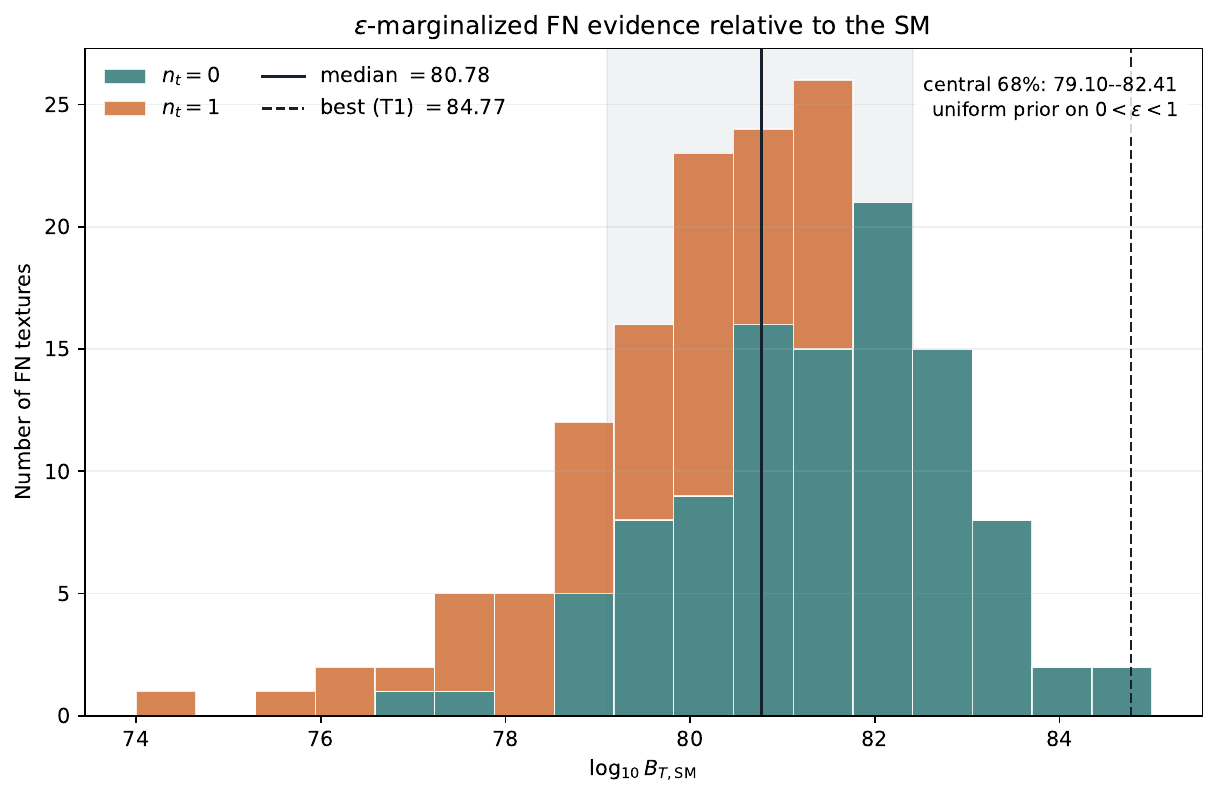}
\caption{Distribution across the 165 FN textures of the base-ten logarithm of
the Bayes factor relative to the SM.  The evidence is marginalized over
$\eps$ using the normalized uniform prior on $0<\eps<1$.  The solid line marks
the median, the shaded region is the central 68\% interval, and the dashed
line marks the best texture T1.  Colors distinguish $n_t=0$ and $n_t=1$.
Each charge assignment is treated as a fixed model, without a charge prior.}
\label{fig:FN-vs-SM-BF-histogram}
\end{figure}

\section{Discussion and outlook}
\label{sec:discussion}

We have studied the naturalness properties of the quark sector of FN models, comparing them to the SM and to one another.
Our analysis also serves as a useful laboratory to test various measures of naturalness, including  several different local sensitivity measures as well as Bayesian evidence.
The local Barbieri-Giudice type metrics are found to have the main drawback that they do not detect when  experimental data is  forcing parameters into regions where
their prior probabilities  are small. In case of the FN models, these are regions with proto-Yukawa couplings noticeably deviating from order one. The standard local
sensitivity measures do not receive any penalty when averaged with the posterior distribution, and hence often lead to too optimistic characterizations
(false negatives).
 However, this conclusion hinges on the type of metric used to measure distances in parameter space. When using a prior-uniform metric (a Euclidean metric in
 transformed coordinates in which priors are flat) the small-prior probability regions do become penalized even for the local measures.
 The Bayesian evidence gives a complementary, explicitly prior-dependent measure of naturalness.
 Its main drawback is that it is only a relative measure between two given models (in the form of Bayes factors). Moreover, its strong prior dependence
 implies that one has to have a very clear idea what the prior should model, in particular what one considers order-one numbers.
 Needless to say, there is an unavoidable degree of subjectivity associated to this assessment.

 The comparison of individual FN textures to the vanilla SM quark sector leads to the unambiguous and expected conclusions of a vast improvement of naturalness.
Even though the low prior probability is missed by  the local logarithmic and linear measures, they still  imply a large amount of tuning over the posterior distributions.
The reason is that there do exist generic  posterior points with order-one Yukawa couplings, but they clearly require fine-tuned cancellations in order to reproduce the data
(i.e., near singular Yukawa matrices).
The BF between the SM and any FN texture takes into account the order-one condition encoded in the priors much more radically, leaving the SM disfavored by roughly 74--85 orders of magnitude across the textures considered here.

There are several directions how this work could be extended.
Firstly, we could include the analysis of the lepton sector.
The charged lepton sector has a mild coupling to  the quark sector via the $\epsilon$ parameter, but it does not seem worthwhile to consider them together, as the number of possible charge assignments grows significantly.
The quark sector by itself is much more constraining than the charged lepton sector because of the CKM matrix. The full lepton sector also requires some further model building to include neutrino masses.
Supersymmetry is another possible direction. The main new ingredient in the quark sector is $\tan\beta$. The  textures studied in this work can be recast as supersymmetric models by interpreting the factor
$\epsilon^{Q_{d_3}}$ as $\tan\beta$, but strictly speaking it is another free parameter.
Lastly, we can study SU(5) or SO(10) compatible charge assignments, possibly within supersymmetry. This links the quark and lepton sectors much more strongly.

\section*{Acknowledgements}
GG acknowledges financial support by the Conselho Nacional de Desenvolvimento
Científico e Tecnológico (CNPq) under fellowship number 313238/2023-5, as well as
the Funda\char"00E7ão de Amparo à Pesquisa do Estado do Rio de Janeiro (FAPERJ) under
project number 210.785/2024. LM is supported by Coordena\char"00E7ão de Aperfei\char"00E7oamento de Pessoal de Nível Superior (CAPES).

\appendix

\section{Priors and metrics for complex parameters}

\label{sec:complex}

For a complex parameter, one typically wants to impose a prior $p(\t)$ that is invariant under phase shifts, i.e., only depends on $|\t|$ and not on $\arg\t$.
This is e.g.~the case for a  Gaussian prior $p(\t)=e^{-\frac{1}{2}|\t|^2}$, which is just the product of Gaussian priors for real and imaginary parts. In general, this simple factorization  is not compatible with phase-shift invariance.
For instance, for a flat prior one should specify limits that only depend on $|\t|$.
Letting $\t=x+iy=re^{i\phi}$
we should, e.g., choose
 $r<1$, and not $-1<x<1$ and $-1<y<1$.
Another example is a log Gaussian prior.  The product of two log Gaussian priors
for real and imaginary parts,
$e^{-\frac{1}{2}\log^2 |x|-\frac{1}{2}\log^2 |y|}d\log|x|\,d\log|y|$
is clearly not rotationally symmetric. The more sensible choice is therefore
$e^{-\frac{1}{2}\log^2 r}d\log r\, d\phi$.

Similar considerations apply to the metric in parameter space.
In the following, we will consider $d_P/2$ complex parameters, but leave all sums over the  indices implicit for clarity.
The linear metric is unambiguous,
\be
ds^2_{\rm lin}=dx^2+dy^2=dr^2+r^2 d\phi^2\,.
\ee
Its volume form is the flat measure $dxdy=rdrd\phi$.
For the log Euclidean metric we have in principle two inequivalent choices
\be
\begin{aligned}
ds^2_{\rm log}={}&r^{-2}dr^2+d\phi^2\,,\\
ds'^2_{\rm log}={}&x^{-2}dx^2+y^{-2}dy^2\,.
\end{aligned}
\ee
Only the former is  invariant under phase shifts, and this is the metric we will use in the definition of
our log sensitivity parameters.
Finally, we need to generalize the prior-uniform metric. We will choose
\be
ds^2_{\rm puc}=d \tilde q_r^2+d\phi^2=\tilde p_r^{2}dr^2+d\phi^2\,.
\ee
Its volume form, $
d{\rm vol}_{\rm puc}=\tilde p_r drd\phi$
is, up to a constant, the prior measure.
For a better comparison with the other fine-tuning metrics, we normalize $ds^2_{\rm puc}$ such that near $r=1$ it coincides with both
the $ds^2_{\rm lin}$ and $ds^2_{\rm log}$.
For a unit Gaussian prior for $\t$, $p_r$ is a Rayleigh distribution and we get
\be
ds^2_{\rm puc}=r^2e^{-r^2+1}dr^2+d\phi^2\,.
\ee
In summary, we have the following sensitivity matrices squared:
\be
\begin{aligned}
S^{}_{\rm lin}S_{\rm lin}\tra={}&S_r S_r\tra+S_{\phi}^{}\frac{1}{r^2}S_{\phi}\tra\,,\\
S^{}_{\rm log}S_{\rm log}\tra={}&S_r r^2S_r\tra+S_{\phi}^{}S_{\phi}\tra\,,\\
S^{}_{\rm puc}S_{\rm puc}\tra={}&S_r \frac{e^{r^2-1}}{r^2}S_r\tra+S_{\phi}^{}S_{\phi}\tra\,,
\end{aligned}
\ee
where $S_r\equiv \partial\log\mu/\partial r$ and $S_\phi\equiv \partial\log\mu/\partial \phi$.

\section{Implementation notes}

We sample the  posterior directly on the best-fit manifold $M$.
In the SM this does not require a Markov chain.  According to
eqs.~(\ref{eq:MSM}) and~(\ref{eq:SMyukawaorbit}), we draw $U_q$, $U_u$ and
$U_d$ independently from the invariant measure on $U(3)$.  Their common
$U(1)$ phase drops out.  Numerically, a Haar-distributed unitary matrix can be sampled easily without a dedicated MC algorithm \cite{Mezzadri:2006Haar}.

For fixed $Q$ and $\eps$, the FN target density is instead
eq.~(\ref{eq:FNposterior}).  We sample it with a Metropolis-within-Gibbs
chain \cite{Roberts:2009Adaptive,Tierney1994} on $(U_q,U_u,U_d)$.  To describe the algorithm, divide the three
generations in each sector $f=q,u,d$ into blocks of equal charge, with block
sizes $n_{f,a}$.  The potential in eq.~(\ref{eq:FNpotential}) is invariant
under the block-diagonal transformations
\be
 U_f\longrightarrow W_fU_f\,,
 \qquad W_f\in\prod_a {\rm U}(n_{f,a})\,,
\ee
i.e., has exact flat directions along the translations $W_f$.
On  $M$, these give
\be
d_{\rm flat}=\sum_{f,a}n_{f,a}^2-1\,,
\qquad
d_{\rm act}=26-d_{\rm flat}
\ee
flat and active directions, respectively. Notice that $8\leq d_{\rm flat}\leq 26$, depending on the amount of degeneracies in the charge assignments.  At the beginning of every sweep
the flat blocks are refreshed by independent rotations drawn from the Haar measure on $\prod_a \text{U}(n_{f,a})$.  This is an
exact update and has acceptance probability one.

The active directions are the "broken" generators $T^A_{f,ij}$, for each pair $(i,j)$ with $Q_{f_i}\neq Q_{f_j}$, and $A=1,2$.
The potential is strongly anisotropic due to the hierarchy parameters.
We visit each of them
once per sweep, in random order, and propose
\be
 U_f\longrightarrow e^{i\delta_{ij} T_{f,ij}^A}U_f\,,
 \qquad
 \delta_{ij}\sim {\cal N}\left(0,\frac{s^2}{m_{f,ij}}\right)\,,
 \qquad
 m_{f,ij}\propto\eps^{-2|Q_{f_i}-Q_{f_j}|}\,.
\ee
The overall normalization of the $m_{f,ij}$ is chosen such that their
geometric mean is one.  This compensates the strong anisotropy of
$V_{Q,\eps}$, while leaving a single step parameter $s$.  The proposal is
accepted with probability $\min(1,e^{-\Delta V})$.  Initially, we tune
$s$ to an average acceptance probability of $0.44$, after which it is kept
fixed.

For every texture we use four independent chains and retain $10^4$ complete
sweeps from each chain.
Equilibration is checked by comparing the
individual chains and the two halves of every chain.  Decorrelation is
monitored through the integrated autocorrelation time $\tau_{\rm int}$, yielding an effective sample size
$N_{\rm eff}=N/\tau_{\rm int}$, using the potential and the invariant block
observables $\|P_{q,a}\hat Y_xP_{x,b}\|^2$, where $\|A\|^2=\tr AA^\dagger$, in addition to the quantities
reported in the main text.  Here $x=u,d$, and $P_{f,a}$ is the diagonal
projector onto the generations in the $a$th equal-charge block of the sector
$f=q,u,d$.

We finally describe the determination of $\eps_\star$ and the evidence.
We use $t=\log\eps$ and define the logarithm of the fixed-$\eps$ evidence,
up to an irrelevant constant $C$ that drops out of all Bayes factors, as
\be
 \ell_Q(t)\equiv\log\left[e^{-tQ_{\rm tot}}
 \int_M e^{-V_{Q,e^t}(u)}dH(u)
 \right]
=\log [p_Q(\mu_{\rm exp}|\eps)/C] \,.
\ee
Its derivative
$\dot \ell_Q\equiv d\ell_Q/dt$ can be written as a posterior expectation value and is
therefore evaluated with the Markov chains described above.  We calculated
$\dot \ell_Q$ this way at several values of $\eps$ around its zero and interpolate it linearly
in $t$.  The solution $\dot \ell_Q(\log\eps_\star)=0$ determines the maximum of the
fixed-$\eps$ evidence.  Integrating the interpolated derivative then gives
the shape $\ell_Q(t)-\ell_Q(\log\eps_\star)$ or equivalently $p_Q(\mu_{\rm exp}|\eps)/p_Q(\mu_{\rm exp}|\eps_\star)$ over the range relevant for the
$\eps$ integral.
The remaining normalization is
\be
p_Q(\mu_{\rm exp}|\eps_\star)=C\eps_\star^{-Q_{\rm tot}}\int_M e^{- V_{Q,\eps_\star}(u)}dH(u) \,.
\ee
It is obtained by sequential Monte
Carlo (SMC) integration \cite{DelMoral:2006SMC}.
Combining the normalization with the fitted shape above gives the full
$p_Q(\mu_{\rm exp}|\eps)$.  The evidence used for the texture
ranking is finally
\be
 p_Q(\mu_{\rm exp})=\int_0^1 p_Q(\mu_{\rm exp}|\eps)\,d\eps,
\ee
corresponding to the uniform prior $p(\eps)=1$ on this interval.  The Bayes
factors are ratios of these integrals, so that the common normalization of
the data density cancels.  The quoted numerical errors include the spread of
the independent SMC determinations and the uncertainty of the interpolated
$\eps$ profile.

\bibliography{paper}

\bibliographystyle{JHEP}

\end{document}